\documentclass[pra,aps,twocolumn,showpacs]{revtex4}
\usepackage{graphicx}
\usepackage{times}
\usepackage{color}
\usepackage{amsmath}
\input epsf

\begin{document}
\title[]
{Theoretical study of pulse delay effects in the photoelectron angular
distribution of near-threshold EUV+IR two-photon ionization  of atoms}

\author{Kenichi L. Ishikawa}
\email[Electronic address: ]{ishiken@n.t.u-tokyo.ac.jp}
\affiliation{Department of Nuclear Engineering and Management, Graduate School of
Engineering, The University of Tokyo, 7-3-1 Hongo, Bunkyo-ku, Tokyo 113-8656, Japan}
\affiliation{Photon Science Center, Graduate School of Engineering, The University of
Tokyo, 7-3-1 Hongo, Bunkyo-ku, Tokyo 113-8656, Japan}

\author{A.~K.~Kazansky}
\affiliation{Departamento de Fisica de Materiales, UPV/EHU,
E-20018 San Sebastian/Donostia, Spain}
\affiliation{Donostia International Physics Center (DIPC),
 E-20018 San Sebastian/Donostia, Spain}
\affiliation{IKERBASQUE, Basque Foundation for
 Science, E-48011 Bilbao, Spain}

\author{N.~M.~Kabachnik}
\affiliation{Skobeltsyn Institute of Nuclear Physics,
 Lomonosov Moscow State University, Moscow 119991, Russia}
\affiliation{European XFEL, D-22761 Hamburg, Germany}

\author{Kiyoshi Ueda}
\affiliation{Institute of Multidisciplinary Research for Advanced Materials,
Tohoku University, Katahira 2-1-1, Aoba-ku, Sendai 980-8577, Japan}

\date{\today}

\begin{abstract}
We theoretically study the photoelectron angular distributions (PADs) from two-color
two-photon
near-threshold ionization of hydrogen and noble gas (He, Ne, and Ar) atoms by a combined action of
femtosecond extreme ultraviolet (EUV) and near-infrared (IR) laser pulses. 
Using the second-order time-dependent perturbation theory, we clarify how the two-photon ionization process depends on EUV-IR pulse delay and how it is connected to the interplay between resonant and nonresonant ionization paths. Furthermore, by
solving the time-dependent Schr\"odinger equation, we calculate the anisotropy parameters $\beta_2$ and
$\beta_4$ as well as the amplitude ratio and relative phase between partial waves characterizing the PADs. 
We show that in
general these parameters notably depend on the time delay between
the EUV and IR pulses, except for He. This dependence is related to the varying relative
role of resonant and nonresonant paths of photoionization.
Our numerical results for H, He, Ne, and Ar show that the pulse-delay effect is more pronounced
for $p$-shell ionization than for $s$-shell ionization.

\end{abstract}

\maketitle

\section{Introduction}\label{sec:intro}

Investigations of non-linear (multiphoton) processes in extreme
ultraviolet (EUV) and soft x-ray energy range is a quickly
developing branch of photon-matter interaction studies. It has been
strongly stimulated by the construction and operation of
EUV and x-ray free electron lasers (FELs) as well as by the progress
in powerful laser physics which results in creation of photon
sources based on high-harmonic generation. Intense
ultra-short photon pulses from FEL allows one to investigate the
non-linear EUV processes using well developed methods of
photoelectron spectroscopy including measurements of photoelectron
angular distributions (PADs), which proved to be a sensitive tool
for studying the dynamics of photoprocesses \cite{Becker96}.

One of the most basic non-linear processes is a two-photon single ionization
(TPSI) of atoms where an atomic electron is emitted by a simultaneous
absorption of two photons. The TPSI (and multi photon ionization more generally) has been intensively
investigated theoretically for decades (see e.g.\cite{Gontier1971PRA,Beers1975PRA,Crance1977PRA,Andrews1977JPB,McClean1978JPB,McClean1979JPB, Jackson1982JPB, Dixit1983PRA, Proulx93,Saenz99,Nikolopoulos01, He2002PRA, Hart05,Selstoe2007PRA,Varma2009PRA,Ishikawa10}) as well as experimentally since the advent of high-harmonic
sources and FELs
\cite{Koboyashi98,Sekikawa04,Miyamoto04,Hasegawa05,Sorokin07,Richardson10,Moshammer11}.
Both the absolute cross section of TPSI \cite{Sato11} and the
angular distribution of photoelectrons \cite{Ma13} have been
recently measured for He.

One of the fundamental problems of TPSI is a relative contribution
of resonant and non-resonant (direct) ionization mechanisms.
Absorption of two photons involves intermediate states of the
system. In general, according to the rules of quantum mechanics, one
should take into account contributions of all excited intermediate
states, both discrete and continuous. In the resonance-enhanced case, i.e., if the photon energy
spectrum allows resonant excitation of one or more excited states, the resonant ionization process via resonant levels and the nonresonant process via nonresonant intermediate levels coexist \cite{Beers1975PRA,Ishikawa12,Ishikawa13}. For a sufficiently long pulse resonant with an excited level, the contribution from the resonant process is dominant, and the TPSI cross section can be calculated
within the two-step approach: excitation of the resonant state and
its subsequent ionization. If we use an ultrashort (femtosecond) exciting pulse with a large bandwidth, on the other hand, the co-presence of resonant and nonresonant contributions becomes a more complex problem.

It has recently been theoretically demonstrated
\cite{Ishikawa12,Ishikawa13} that the angular distribution of
photoelectrons in TPSI generated by ultrashort EUV pulses changes with the pulse width, reflecting the competition between
resonant and nonresonant ionization paths. Calculations for
H and He atoms have shown that the relative phase $\delta$
between $S$ and $D$ ionization channels is distinct from the
scattering  phase shift difference and varies with the pulse width
and that this variation is different for different photon energies. This
prediction has been confirmed experimentally \cite{Ma13} for the case of He.

The above discussion, which originally concerns single-color TPSI, is quite general and can also be extended to two-color cases. Specifically let us consider a combined action of an EUV pulse from FEL or
high-harmonic source and of a synchronized optical laser pulse.
Such two-color multi-photon ionization experiments have proved to be useful for characterization of ultrashort EUV pulses as well as for a detailed investigation of ionization
dynamics \cite{Glover96,Toma00,Meyer06,Meyer08,Fushitani13} (see
also review \cite{Meyer10}). Here also the measurements of PADs
provided deeper insight into the physics of the photon-atom
interaction
\cite{Guyetand05,Guyetand08,Haber09a,Haber09b,Haber10,Haber11,Keeffe13,Mondal13}.
One of the advantages of the two-color investigation is that the
two pulses can be independently controlled; one has possibility to vary the
frequency, duration and polarization of the EUV and
optical pulses independently. This gives much more flexibility to the experiment.
One additional advantage is that in two-color experiments with
ultrashort pulses one can study the time-evolution of the ionization
process by controlling the time delay between them.

Recently this additional degree of freedom has been used to
advantage in Ref. \cite{Mondal14}. It was shown experimentally that
the PAD in two-color TPSI of Ne atoms is notably different for
temporally overlapping and non-overlapping EUV and IR
pulses. The difference between these two extreme cases clearly
demonstrates that the PAD in TPSI strongly depends on the time delay
between the pulses. The corresponding theoretical calculations agree with the measurements and explain the dependence by the change in the relative contribution of resonant and nonresonant ionization paths\cite{Mondal14}.

In the present work we extend our previous works \cite{Ma13,Ishikawa12,Ishikawa13,Mondal14} and theoretically investigate in more detail the pulse-delay dependence of photoelectron energy spectra and angular distributions, both energy-resolved and integrated, in near-threshold two-color TPSI, with focus on 
resonant and nonresonant contributions. We first describe two-color TPSI with the second-order time-dependent perturbation theory and show how the interplay between the resonant and nonresonant paths depends on the pulse delay. Then, we study the pulse-delay effect for different target atoms (H, He, Ne, and Ar), based on direct numerical simulation of the time-dependent Schr\"odinger equation (TDSE).

This paper is organized as follows. In the next section, using the perturbation theory, 
we discuss the general idea of the relation between the pulse-delay dependence of the final-state 
amplitude for two-color TPSI and the contribution of
resonant and nonresonant ionization paths. We then shortly
describe the numerical methods used to solve TDSE and calculate anisotropy parameters. In Sec.
\ref{sec:results} we present and discuss the simulation results for H, He, Ne and Ar atoms. 
Our conclusions and
outlook are presented in Sec. \ref{sec:conclusions}.

\section{Analysis based on the perturbation theory}
\label{sec:theor}


To illustrate the main idea it is instructive to consider the
problem of TPSI within the second-order time-dependent perturbation
theory. Generalizing the expression for the amplitude of the
two-photon transition presented in Ref. \cite{Dudovich01} to the
case of multiple intermediate levels, we can write the amplitude of
the final state $f$ of the atom as (atomic units are used throughout
unless otherwise indicated)
\begin{multline}
 c_f= i\sum_\alpha\mu_{f\alpha}\mu_{\alpha
i}\left[i\pi \hat{E}(\omega_{\alpha i}) \hat{E}(\omega_{f\alpha})\right.\\
+\left. P\int_{-\infty}^{\infty}\frac{\hat{E}(\omega)\hat{E}(\omega_{fi}-\omega)}
{\omega_{\alpha i}-\omega}d\omega
 \right],
 \label{eq:cfgeneral}
\end{multline}
where $\mu_{\alpha i}$ etc. denote the dipole transition matrix
element between state $i$ and $\alpha$; $i$ is the initial state,
$\alpha$ the intermediate states ($\alpha$ should be taken as a
collection of quantum numbers that specify each energy eigenstate,
e.g., $\alpha=(n,l,m)$ for bound states and $\alpha=(\epsilon,l,m)$
for continuum states for the case of a hydrogen-like atom);
$\omega_{\alpha i}=\omega_\alpha-\omega_i$ etc., $P$ is the Cauchy
principal value, and $\hat{E}(\omega)$ the Fourier transform of the
electric field $E(t)$ of the ionizing pulse. In principle, the sum
should be taken over all the bound and continuum intermediate states
$\alpha$. The first and second terms of Eq.\ (\ref{eq:cfgeneral})
can be interpreted as the resonant (or two-step) and non-resonant
processes, respectively.

Let us consider a double (EUV + IR) pulse of the form:
\begin{equation}
E(t) = E_X(t) + E_{IR}(t-\tau),
\end{equation}
and its Fourier transform:
\begin{equation}
\hat{E}(\omega)=\hat{E}_X(\omega)+\hat{E}_{IR}(\omega)e^{i\omega\tau},
\end{equation}
where the first and the second terms correspond to the EUV and IR
pulses, respectively, and $\tau$ denotes the delay between the
pulses. In this case, neglecting resonant excitation from the ground
state by an IR photon [i.e., $\hat{E}_{IR}(\omega_{\alpha i})\approx
0$ for any $\alpha$], Eq.\ (\ref{eq:cfgeneral}) can be approximated
by
\begin{widetext}
\begin{eqnarray}
c_f&=& i\sum_\alpha\mu_{f\alpha}\mu_{\alpha i}\left[i\pi
\hat{E}_{X}(\omega_{\alpha i}) \hat{E}_{IR}(\omega_{f\alpha})
e^{i\omega_{f\alpha}\tau}\right.\nonumber\\
\label{eq:cfdoublepulse} &+& \left.
P\int_{-\infty}^{\infty}\frac{\hat{E}_X(\omega_{fi}-\omega)
\hat{E}_{IR}(\omega)e^{i\omega\tau}
+\hat{E}_X(\omega)\hat{E}_{IR}(\omega_{fi}-\omega)e^{i(\omega_{fi}-\omega)\tau}
}{\omega_{\alpha i}-\omega}d\omega
 \right] \\
&=& i\sum_\alpha\mu_{f\alpha}\mu_{\alpha i}\left[i\pi
\hat{E}_{X}(\omega_{\alpha i}) \hat{E}_{IR}(\omega_{f\alpha})
e^{i\omega_{f\alpha}\tau}
\label{eq:cfdoublepulse2} 
+P\int_{-\infty}^{\infty}\hat{E}_X(\omega_{fi}-\omega)\hat{E}_{IR}
(\omega)e^{i\omega\tau}\left(\frac{1}{\omega_{\alpha
i}-\omega}-\frac{1}{\omega_{f \alpha}-\omega}\right)d\omega
 \right]\,.
\end{eqnarray}

When the two pulses overlap ($\tau=0$),
\begin{equation}
c_f = i\sum_\alpha\mu_{f\alpha}\mu_{\alpha
i}\left[i\pi \hat{E}_{X}(\omega_{\alpha i})
\hat{E}_{IR}(\omega_{f\alpha}) +  P\int_{-\infty}^{\infty}\hat{E}_X(\omega_{fi}-\omega)\hat{E}_{IR}(\omega)
 \left(\frac{1}{\omega_{\alpha i}-\omega}-\frac{1}{\omega_{f
\alpha}-\omega}\right)d\omega
 \right],
\label{eq:overlap} 
\end{equation}
\end{widetext}
thus, both the first (resonant) and second (non-resonant) terms
contribute to $c_f$, leading to an additional phase and to a
photoelectron angular distribution (PAD) different from the one
expected from the scattering phase shifts. With increasing delay,
factors $e^{i\omega_{f\alpha}\tau}$ and $e^{i\omega\tau}$ begin to
oscillate, and the PAD changes with $\tau$.

For large delay, Eq.\ (\ref{eq:cfdoublepulse2}) can be transformed,
after some algebra, into
\begin{multline}
c_f = -\pi \\
\times \sum_\alpha\mu_{f\alpha}\mu_{\alpha i}[1+sgn(\tau)]
\hat{E}_{IR}(\omega_{f\alpha})\hat{E}_{X}(\omega_{\alpha
i})e^{i\omega_{f\alpha}\tau}\,, 
\label{eq:separate}
\end{multline}
where we have used the relation
\begin{equation}
P\int_{-\infty}^{\infty}\frac{e^{i\omega\tau}}{\omega_0-\omega}d\omega
= -i\pi e^{i\omega_0\tau} sgn(\tau).
\end{equation}
If the EUV spectrum is located within the Rydberg manifold, a
Rydberg wave packet is formed by the EUV pulse and then ionized by
the IR pulse. The factor $e^{i\omega_{f\alpha}\tau}$ describes the
evolution of the Rydberg wave packet with increasing delay $\tau$.
The ionization yield $|c_f|^2$ changes with $\tau$, reflecting the
Kepler-like motion of the Rydberg wave packet, while the PAD only
slightly changes (nearly constant) with $\tau$. There is no
ionization ($c_f = 0$) if the IR pulse precedes the EUV pulse
($\tau<0$), as is evident if one considers in the time domain.
Apparently, Eq.\ (\ref{eq:separate}) indicates that there are only
resonant paths, which might sound obvious again in the time-domain
consideration. It should be, however, noticed that the second term
in the sum in Eq.\ (\ref{eq:separate}) originates from the second
term in Eq.\ (\ref{eq:overlap}), usually interpreted as non-resonant
paths. This observation implies that the attribution of resonant and
non-resonant processes may be somewhat arbitrary.


In the above-threshold case, where the EUV spectrum lies above the
ionization threshold, the two-photon ionization yield vanishes if
the two-pulses are separated in time. This intuitively obvious result can be shown as follows.
Assuming that the transition matrix elements in Eq.\
(\ref{eq:separate}) are almost constant within the bandwidth of the
pulses, one finds,
\begin{equation}
c_f \propto \int_{-\infty}^\infty
\hat{E}_{IR}(\omega_{f\alpha})\hat{E}_{X}(\omega_{\alpha
i})e^{i\omega_{f\alpha}\tau}d\omega_{\alpha}.
\end{equation}
After some algebra using Parseval's theorem:
\begin{equation}
\label{eq:Parseval} \int_{-\infty}^\infty
\hat{f}(\omega)\hat{g}(\omega)d\omega = \int_{-\infty}^\infty
f(t)g(-t)dt,
\end{equation}
one obtains,
\begin{multline}
\int_{-\infty}^\infty
\hat{E}_{IR}(\omega_{f\alpha})\hat{E}_{X}(\omega_{\alpha
i})e^{i\omega_{f\alpha}\tau}d\omega_{\alpha} \\
=\int_{-\infty}^\infty
E_{X}(t) E_{IR}(t-\tau)e^{-i\omega_{fi}\tau}dt,
\end{multline}
which vanishes if the XUV and IR pulses do not overlap each other at
all, i.e., $E_{X}(t) E_{IR}(t-\tau) = 0$ for any $t$.

\section{Numerical solution of time-dependent Schr\"odinger equation}\label{sec:TDSE}

In the numerical calculations discussed below, we consider the case
where the EUV photon energy is slightly below 
the ionization threshold. We have chosen the
following basic parameters of the pulses which are rather common in
recent experiments: the IR pulse with the carrier frequency
$\omega_L = 1.55$ eV (800 nm) has a duration of 30 fs (FWHM of
intensity). The peak intensity of the IR field is $10^{10}$ W/cm$^2$
which is sufficiently low to guarantee that only one IR photon is
absorbed in ionization. The duration of the EUV pulse is 8 fs (FWHM of
intensity), typical of a coherent time of an EUV FEL pulse \cite{Moshammer11}. 
The time delay between maxima of
the pulses is varied from 0 (complete overlap of the pulses)  to 160
fs at most. The EUV photon energy $\hbar\omega_X$ is chosen to be
by 0.2 eV smaller than the ionization potential $I_p$ of each atom, i.e., 
the excess energy $E_{ex}\equiv \hbar\omega_X - I_p$ is -0.2 eV.

We assume that both the EUV and IR pulses are
linearly polarized along the $z$ direction. 
The photoelectron angular distribution from two-photon
ionization is given by \cite{Smith1988AAMP},
\begin{equation}
\label{eq:pad}
I(\theta)=\frac{\sigma}{4\pi}\left[1+\beta_2P_2(\cos\theta)+
\beta_4P_4(\cos\theta)\right],
\end{equation}
where $\sigma$ is the total cross section, $\theta$ is the angle
between the laser polarization and the electron velocity vector, and
$\beta_2$ and $\beta_4$ are the anisotropy parameters associated
with the second- and fourth-order Legendre polynomials, $P_2(x)$ and
$P_4(x)$, respectively.

Although in principle it would be possible to calculate PADs using the analytical expression 
given in the previous section, it would be very complicated to perform an integration over all the bound and continuum intermediate states. Instead, it is easier and
more straightforward to numerically solve the time-dependent
Schr\"odinger equation and to obtain from its solution the
amplitudes of photoionization and then cross sections, angular
distributions {\it etc}. The TDSE in the dipole approximation and the length-gauge, describing the
evolution of an atom under the action of two-color pulses is 
written as,
\begin{widetext}
\begin{equation}
\label{eq:TDSE_gen} i\frac{\partial\Phi ({\bf
r_1,...r_n},t)}{\partial t} =\left[ \hat H_e({\bf r_1,...r_n}) -
\sum_i^n z_i (E_X(t) + E_{IR}(t-\tau)) \right]\Phi ({\bf
r_1,...r_n},t)\,,
\end{equation}
\end{widetext}
where $\Phi ({\bf r_1,...r_n},t)$ denotes the wave function of
the atom and $\hat H_e({\bf r_1,...r_n})$ the field-free atomic Hamiltonian.

We exactly solve the TDSE (\ref{eq:TDSE_gen}) for H and He, while we make some additional
approximations for the case of multi-electron atoms (Ne and Ar). Below we briefly summarize 
the numerical methods applied in this paper.

\subsection{Hydrogen atom}
\label{sec:TDSE-H}

For a hydrogen atom the TDSE (\ref{eq:TDSE_gen}) is reduced to, 
\begin{multline}
i\frac{\partial\Phi ({\bf r},t)}{\partial t}\\
=\left[-\frac{1}{2}\nabla^2
-\frac{1}{r}-z(E_X(t)+E_{IR}(t-\tau))\right]\Phi ({\bf r},t).
\label{eq:TDSE_H}
\end{multline}
Equation (\ref{eq:TDSE_H}) is numerically integrated using the
alternating direction implicit (Peaceman-Rachford) method
\cite{Kulander1992,He2002PRA,KLI2002PRA,KLI2003PRL,KLI2006PRA,KLI2007PRA,Schiessl2007PRL,KLI2009PRA,Arbo2010PRA,Ishikawa10}.
Sufficiently long (typically a few times the pulse width) after the
pulse has ended, the ionized wave packet moving outward in time is
spatially well separated and clearly distinguishable from the
non-ionized part remaining around the origin. We calculate the
parameters $\beta_2$ and $\beta_4$ by integrating the ionized part
of $|\Phi ({\bf r})|^2$ over $r$ and $\phi$.

For the case of s-shell ionization (H and He) by two dipole photons,
the angular distribution of photoelectrons is determined by the
interference of the $S$ and $D$ wave packets
\cite{Ishikawa12,Ishikawa13},
\begin{equation}
I(\theta)\propto \left| \tilde{c}_Se^{i\delta_0}Y_{00} -
\tilde{c}_De^{i\delta_2}Y_{20}\right|^2\,,
\end{equation}
where $Y_{00}$ and $Y_{20}$ are spherical functions, $\tilde{c}_S$
and $\tilde{c}_D$ are real numbers that have the same absolute values as complex amplitudes $c_S$ and $c_D$, respectively, and $\delta_l$ the phase of the partial
wave, or the {\it apparent} phase shift. The apparent phase shift
difference,
\begin{equation}
\label{eq:apparent_phase} \delta\equiv
\delta_0-\delta_2=\delta_{sc}+\delta_{ex},
\end{equation}
consists of a part $\delta_{sc}$ intrinsic to the continuum
eigenfunctions (scattering phase shift difference), which has
previously been studied both theoretically
\cite{Oza1986PRA,Chang1995PRA,Gien2002JPB} and experimentally
\cite{Haber09b}, and the extra contribution $\delta_{ex}=\arg
c_S/c_D$ from the competition of the resonant and non-resonant
paths. One obtains the amplitude ratio
$W\equiv\tilde{c}_S/\tilde{c}_D$ and the phase-shift difference
$\delta$ from the anisotropy parameters using the relations
\cite{Ishikawa12,Ishikawa13}
\begin{equation}
\label{eq:beta2and4}
\beta_2=\frac{10}{W^2+1}\left[\frac{1}{7}-\frac{W}{\sqrt{5}}\cos\delta\right],
\quad \beta_4=\frac{18}{7(W^2+1)}.
\end{equation}

It should be noted that the values of $\beta_2$, $\beta_4$, $W$, and
$\delta$ obtained as above are integrated over photoelectron energy.
We calculate, on the other hand, energy-resolved values from $c_S$
and $c_D$ obtained by directly projecting the $S$ and $D$ partial
waves onto the Coulomb wave functions.\\

\subsection{Helium atom}
\label{sec:TDSE-He}

To describe the photoionization of He atom we use direct numerical
solution of the full-dimensional two-electron TDSE in the length gauge \cite{ATDI2005},
\begin{multline}
i\frac{\partial\Phi ({\bf r}_1,{\bf
r}_2,t)}{\partial t} \\
= [H_e+ (z_1+z_2)(E_X(t) + E_{IR}(t-\tau))]\Phi
({\bf r}_1,{\bf r}_2,t),
\label{eq:TDSE_He} 
\end{multline}
with the atomic Hamiltonian,
\begin{equation}
\label{eq:atomic_hamiltonian} H_e = -\frac{1}{2}\nabla_1^2
-\frac{1}{2}\nabla_2^2- \frac{2}{r_1}-\frac{2}{r_2}+\frac{1}{|{\bf
r}_1-{\bf r}_2|}.
\end{equation}
We solve Eq.\ (\ref{eq:TDSE_He}) numerically using the
time-dependent close-coupling method
\cite{ATDI2005,Parker2001JPB,Pindzola1998PRA,Pindzola1998JPB,Colgan2001JPB}.
Similarly to the case of a hydrogen atom, sufficiently long after
the pulse has ended, the ionized wave packet moving outward in time
is spatially well separated and clearly distinguishable from the
non-ionized part remaining around the origin. We calculate
photoelectron-energy-integrated $\beta_2$ and $\beta_4$ by
integrating the ionized part of $|\Phi ({\bf r}_1,{\bf r}_2)|^2$
over $r_1,r_2,\theta_2,\phi_1,\phi_2$, from which one obtains $W$
and $\delta$ by solving Eqs.\ (\ref{eq:beta2and4}). We use the
values of $\delta_{sc}$ from \cite{Gien2002JPB} to calculate
$\delta_{ex}=\delta-\delta_{sc}$.\\

\begin{figure}
\centering
\includegraphics[width=83mm]{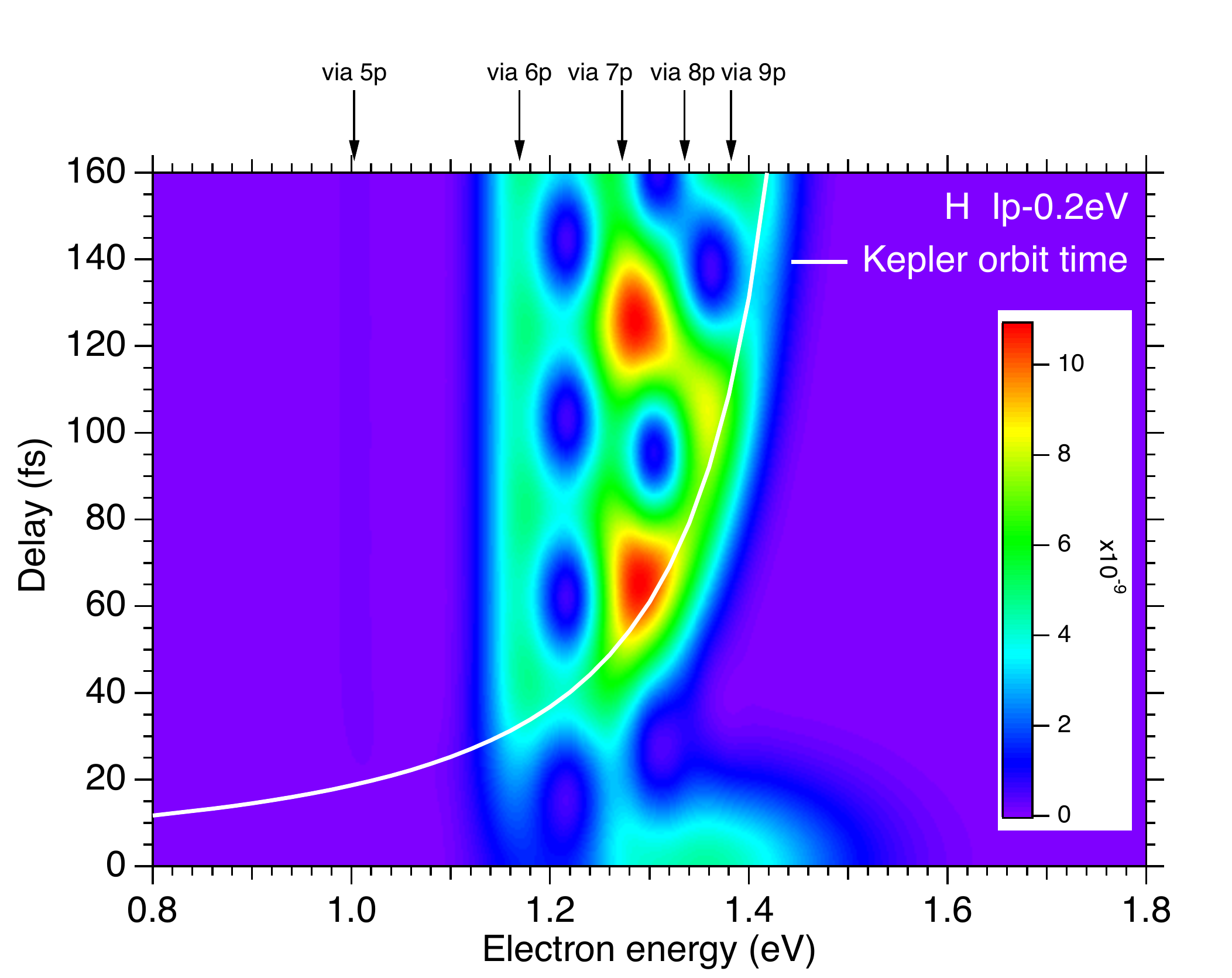}
\caption{(color online) False-color representation of photoelectron  energy spectra
as a function of time delay for the case of H atom. Above the top
axis, the energy positions corresponding to a single IR photon
ionization from each of $5p$-$9p$ levels are indicated with vertical
arrows. White solid line: the Kepler orbit time corresponding to
$E_{\rm kin}-\hbar\omega_L$ (see
text).}\label{fig:photoelectron-energy-spectra-H}
\end{figure}

\subsection{Neon and argon atoms}
\label{sec:TDSE-Ne-Ar}

For multi-electron atoms, like Ne and Ar, direct numerical solution
of Eq. (\ref{eq:TDSE_gen}) is impossible. In many cases it is
sufficient to solve the TDSE for one electron only (single active
electron approximation) ignoring electron-electron correlations and
influence of external electromagnetic fields on the other electrons
\cite{Kazansky06,Kazansky07a,Kazansky07b}. In the present study we
use this approach for two-color photoionization of Ne and Ar. In
contrast to H and He cases, in Ne and Ar atoms the outermost
 ``active" electron has $p$-symmetry and therefore it can be initially
in $p_\sigma \; (m=0)$ and $p_\pi \; (m=1)$ states. Due to axial
symmetry of the problem, ionization of states with $\sigma \; (m=0)$
and $\pi \; (m=1)$ symmetry can be considered independently and then
the obtained cross sections should be summed incoherently.

Since the magnitude of the considered EUV field is comparatively low
and its frequency is high, we use the first order perturbation
treatment and the rotating wave approximation (RWA) for the
description of the EUV-pulse interaction with the atomic $p$-electron.
Thus, we present the active electron wave function as the following
sum,
\begin{equation}\label{eq:function}
\Phi_{pm} ({\bf r},t) = \exp (-i\epsilon_{p}t)\phi^{(0)}_{p m} ({\bf
r})+ \phi_{pm} ({\bf r},t) \,.
\end{equation}
Here $\epsilon_p$ is the binding energy of the electron in the
initial state, $ \phi_{pm} ({\bf r},t)$ describes a perturbation of
the active electron wave function due to interaction with the EUV
field, and  $ \phi^{(0)}_{pm} ({\bf r})$ is the wave function of the
active electron in the initial state. Within the RWA, the TDSE for
an active $p$-electron can be written as,
\begin{multline}
i\frac{\partial\phi_{pm} ({\bf r},t)}{\partial t}\\
=\left[ -\frac{1}{2}\nabla^2 + U(r)  -  z E_{IR}(t-\tau) \right]\phi_{pm}
({\bf r},t) \\
-\frac{1}{2} z \bar E_X(t) \exp(-i(\epsilon_p +
\omega_X)t) \phi^{(0)}_{pm} ({\bf r})\, ,
\label{eq:TDSE_Ne}
\end{multline}
where $\omega_X$ and $\bar E_X(t)$ denote the carrier frequency and the envelope of the EUV pulse,
respectively. The interaction of the active
electron with the ion core is taken into account by the effective
single-electron potential $U(r)$. In the present study for the atoms
Ne and Ar we have used the Herman-Skillman potential obtained within
the Hartree-Slater approximation \cite{HS63}. To solve Eq.\,
(\ref{eq:TDSE_Ne}) we used a method based on the expansion of the
wave packet $\phi_{pm} ({\bf r},t)$ in partial waves. The method is
described in details in Refs. \cite{Kazansky07a,Kazansky07b}. The
calculated double differential cross section was further used for
calculating the asymmetry parameters $\beta_n$ as functions of
photoelectron energy.\\

\section{Results and discussion}\label{sec:results}

\subsection{Show-case of hydrogen atom two-photon near-threshold ionization}\label{sec:results-H}

\begin{figure}
\centering
\includegraphics[width=83mm]{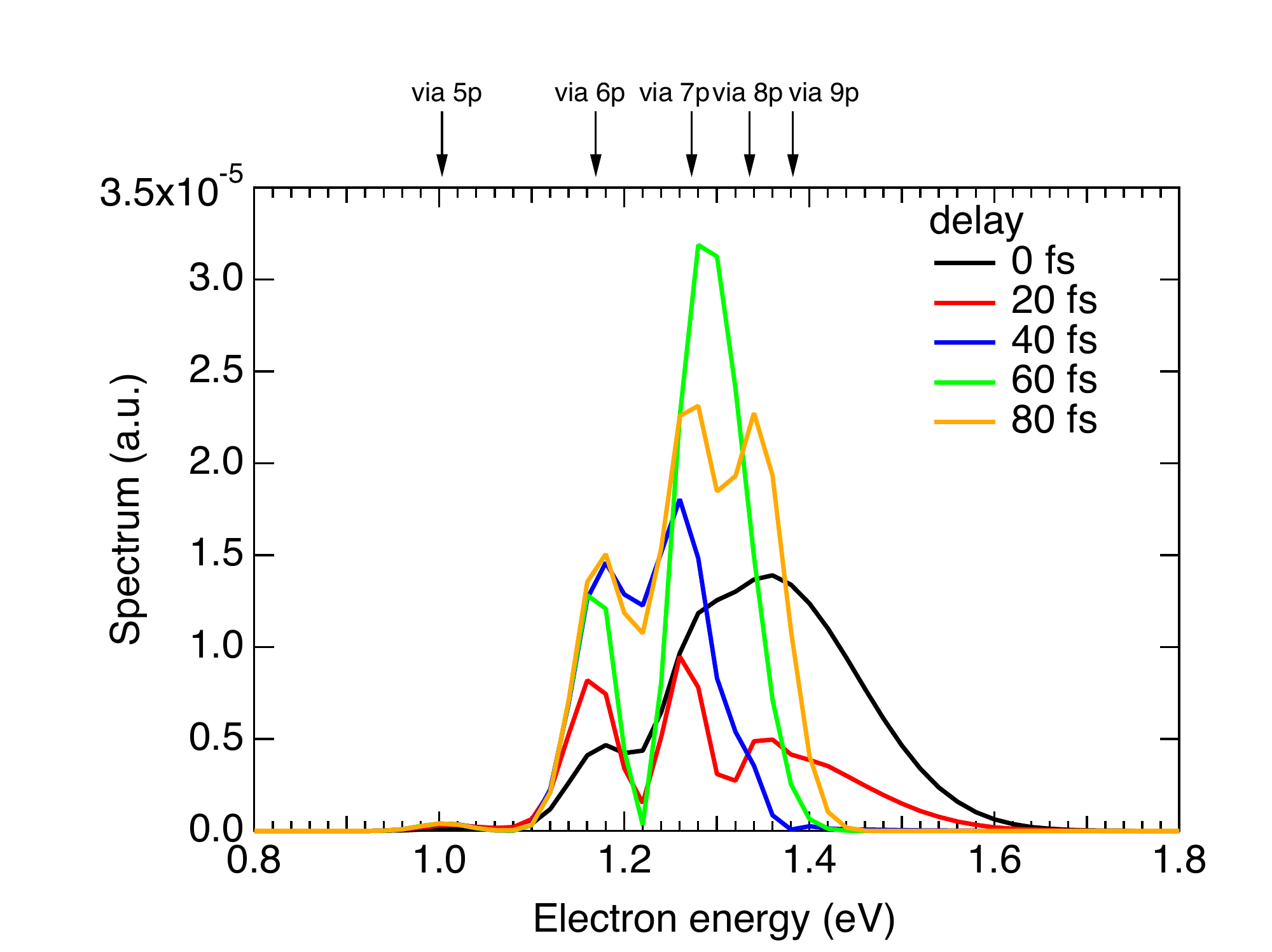}
\caption{(color online) Photoelectron energy spectra for H atom for several values
of time delay indicated in the legend. Above the top axis, the
energy positions corresponding to a single IR photon ionization from
each of $5p$-$9p$ levels are indicated with vertical arrows.}
\label{fig:photoelectron-energy-spectra-H-2}
\end{figure}

In this subsection we discuss in detail TPSI of a hydrogen atom as a
show case demonstrating all peculiarities of the process of the
two-color near-threshold ionization. The parameters of the pulses
are given in Sec.\ \ref{sec:TDSE}. In addition, the peak EUV intensity is set to $10^6\,{\rm W/cm}^2$ (the process under consideration is basically linear in EUV intensity). The time delay between
the pulse peaks is varied from 0 (complete overlap) to 160 fs where
the IR pulse is completely separated from the preceding EUV pulse.

\begin{figure}
\centering
\includegraphics[width=83mm]{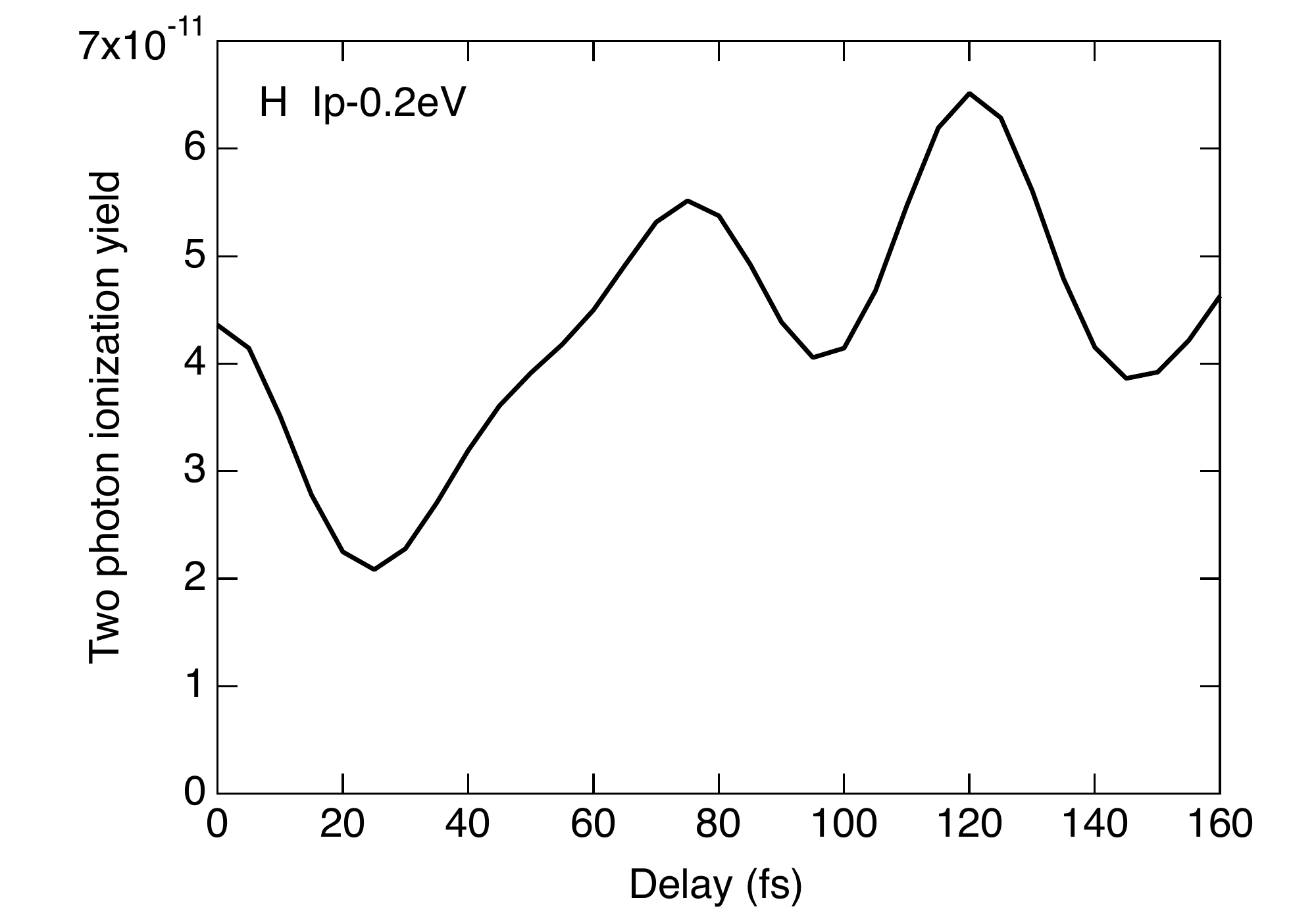}
\caption{Photoelectron yield from H atom as a function of time
delay}\label{fig:TPI-yield-H}
\end{figure}
\begin{figure}
\centering
\includegraphics[width=83mm]{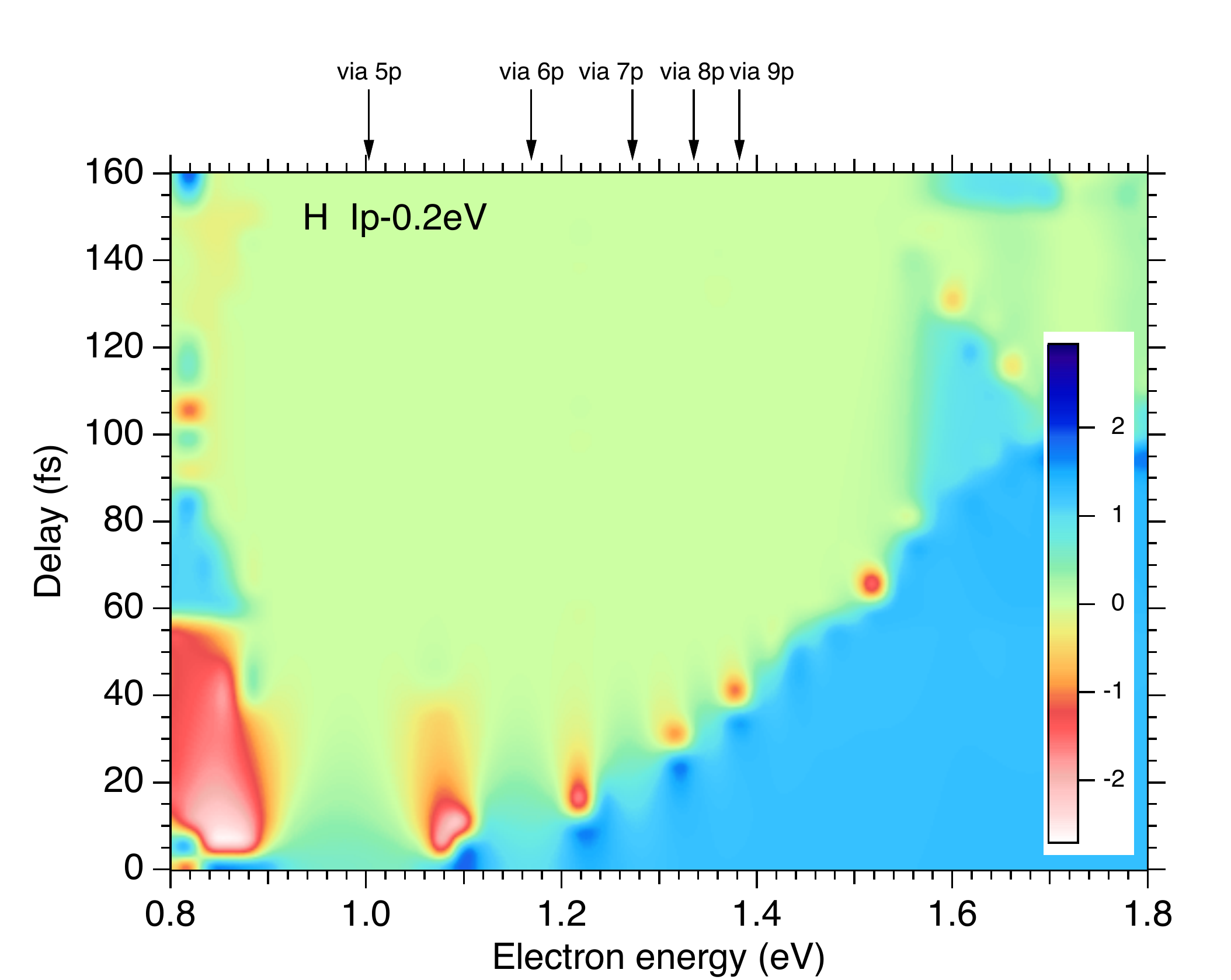}
\caption{(color online) False-color representation of the extra phase  shift
difference $\delta_{ex}$ as a function of time delay and
photoelectron energy for H atom. Above the top axis, the energy
positions corresponding to a single IR photon ionization from each
of $5p$-$9p$ levels are indicated with vertical
arrows.}\label{fig:delta-delay-energy}
\end{figure}

Figure \ref{fig:photoelectron-energy-spectra-H} illustrates how the
photoelectron  energy spectrum varies with the delay between the
pulses in false-color representation. Figure
\ref{fig:photoelectron-energy-spectra-H-2} plots the spectra for
several values of delay from 0 to 80 fs. The results are shown for
the EUV photon energy of 13.405 eV, which is 0.2 eV below threshold
($E_{ex} = -0.2$ eV). In these figures, the kinetic energy positions $E_{\rm
kin}=\omega_L-\frac{1}{2n^2}\,(n=5,\cdots,9)$ corresponding to a
single IR photon ionization from each of $5p$-$9p$ levels are
indicated with vertical arrows. At $\tau = 0$ where the two pulses
overlap each other, resonant peaks are embedded in a broad spectrum
due to non-resonant processes, centered at $E_{\rm
kin}=\hbar\omega_X+\hbar\omega_L-I_p({\rm H}) =
\hbar\omega_L+E_{ex}=1.35\,{\rm eV}$. With increasing delay, the
spectrum is dominated by resonant peaks, and the components between
the peaks exhibit clear interference pattern, related with the
evolution of the Rydberg wave packet created by the EUV pulse. The
white solid line in Fig.\ \ref{fig:photoelectron-energy-spectra-H}
plots the nominal Kepler orbit time,
\begin{equation}
\label{eq:kepler} \tau_n = 2\pi n^3,
\end{equation}
corresponding to a Rydberg state with the principal quantum number
$n$ from  which the photoelectron energy is achieved through an IR
photon absorption, i.e., $E_{\rm kin}=\omega_L-\frac{1}{2n^2}$. One
can see that this line indeed coincides with the first interference
maximum. Due to this Rydberg wave packet dynamics, the two-photon
ionization yield integrated over the photoelectron energy $E_{\rm
kin}$ oscillates with the delay [Fig.\ \ref{fig:TPI-yield-H}].

\begin{figure}
\centering
\includegraphics[width=83mm]{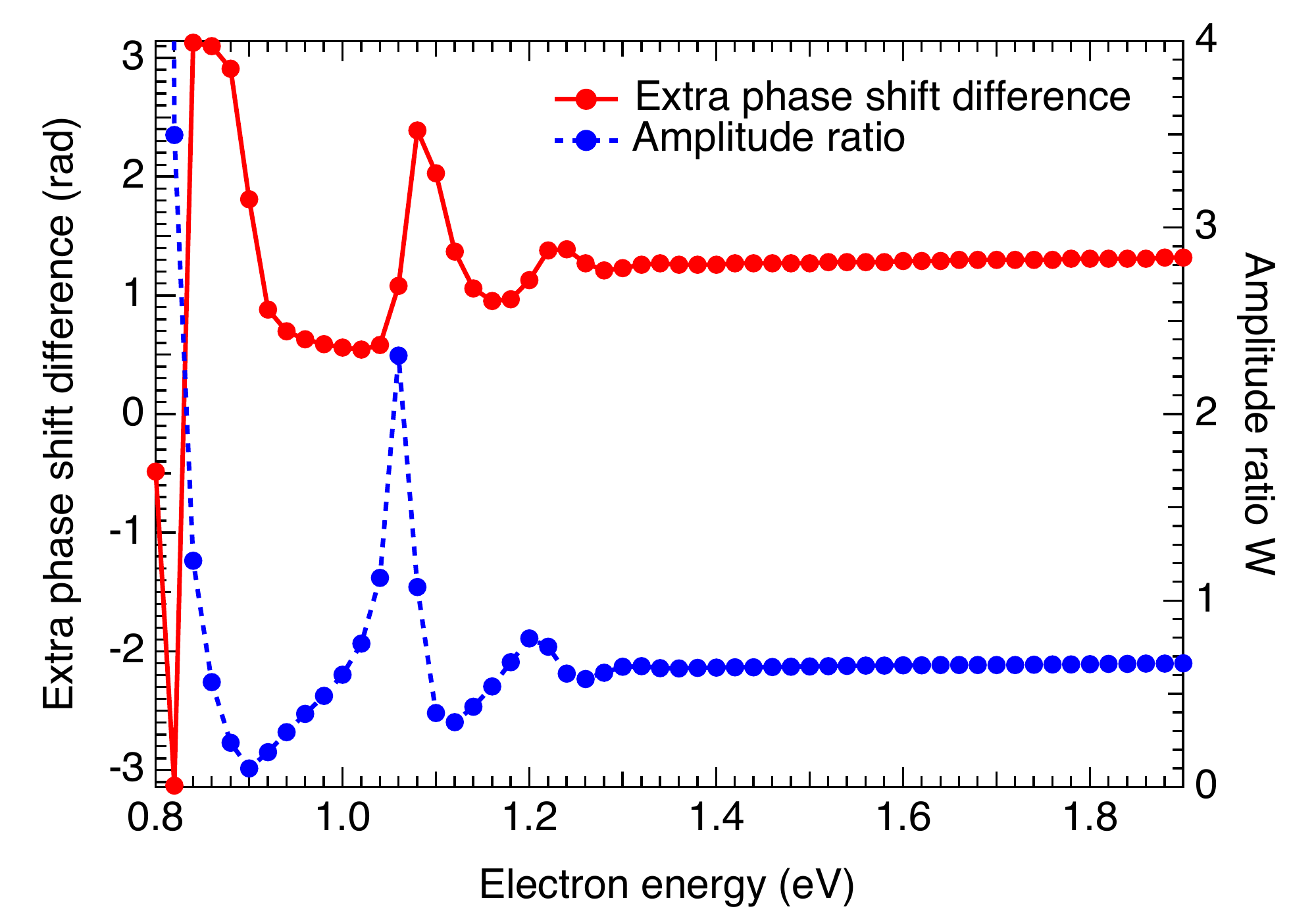}
\caption{(color online) Extra phase shift difference $\delta_{ex}$  (left axis) and
amplitude ratio $W$ (right axis) as a function of photoelectron
energy for $\tau=0$ for H atom.}\label{fig:delta-tau-0}
\end{figure}
\begin{figure}
\centering
\includegraphics[width=83mm]{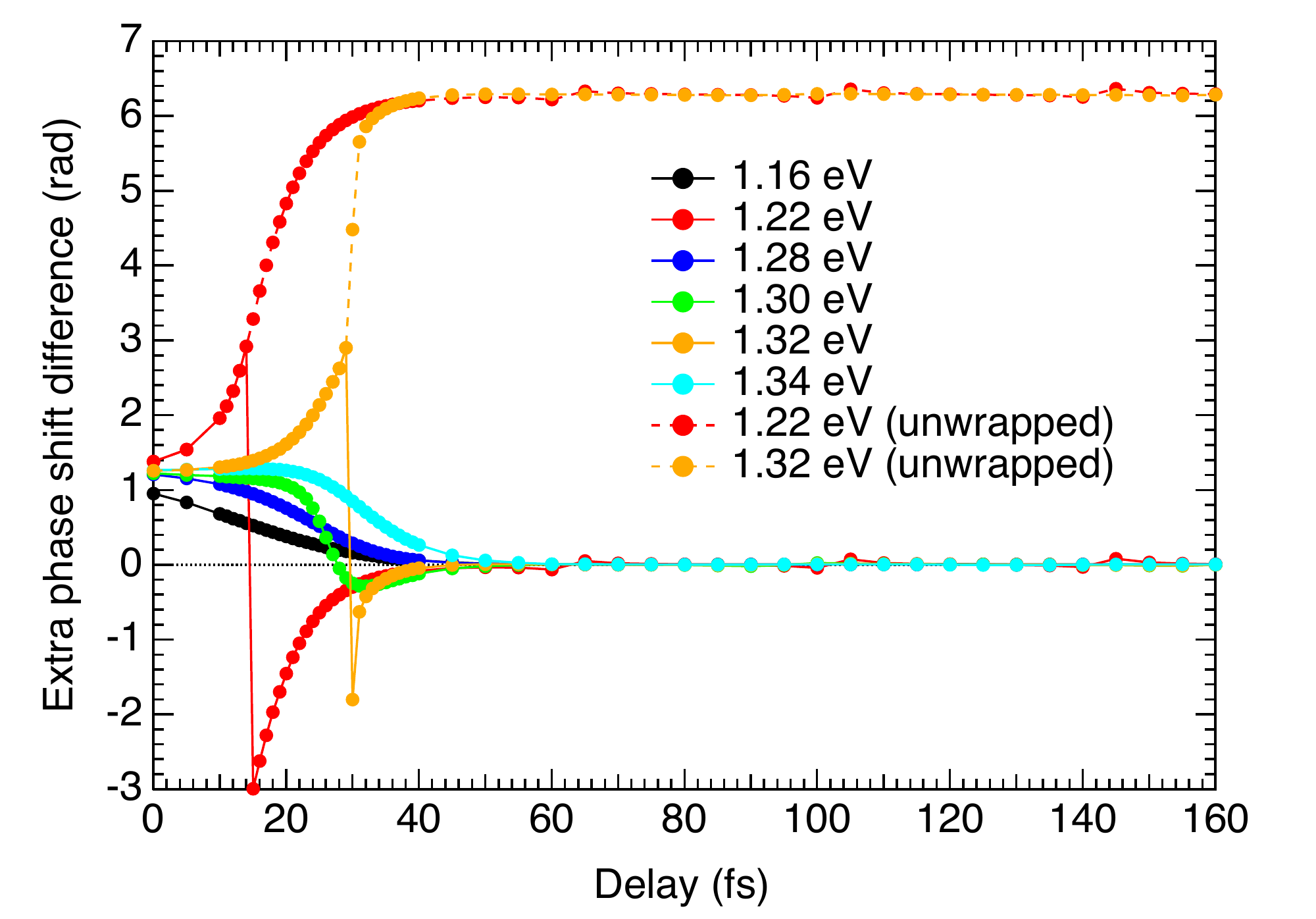}
\caption{(color online) Extra phase shift difference $\delta_{ex}$ as a  function
of time delay for H atom for several photoelectron energies
indicated in the legend. Solid: wrapped within  the range $-\pi \le
\delta_{ex} \le \pi$, dashed (for 1.22 and 1.32 eV): unwrapped with
a modulus of $2\pi$.}\label{fig:delta-delay}
\end{figure}

We show in Fig.\ \ref{fig:delta-delay-energy} the extra phase shift
difference $\delta_{ex}$ as a function of time delay and
photoelectron energy in false-color representation. While
$\delta_{ex}$ is finite at zero delay, it varies with increasing
delay and vanishes at large delay within the energy range
($1.0\,{\rm eV} \lesssim E_{\rm kin} \lesssim 1.5\,{\rm eV}$) of
photoelectrons, as predicted in Sec. \ref{sec:theor}. Figure
\ref{fig:delta-tau-0} plots the dependence of $\delta_{ex}$ and $W$
on photoelectron energy. In the low-energy part, they oscillate,
reflecting changing relative contribution of resonant and
non-resonant paths, whereas they are nearly constant in the
high-energy part ($\gtrsim 1.3\,{\rm eV}$) for which $E_{\rm kin} -
\hbar\omega_L$ lies in the Rydberg manifold whose level spacing is
much smaller than the spectral width.

One can see from Fig. \ref{fig:delta-delay-energy} that the
variation of $\delta_{ex}$ with increasing delay is not necessarily
monotonic. To take a closer look at this, we plot the
delay-dependence of $\delta_{ex}$ for several photoelectron energies
in Fig.\ \ref{fig:delta-delay}. For 1.16, 1.28, and 1.34 eV with a
single dominant intermediate state ($6p, 7p$, and $8p$,
respectively), $\delta_{ex}$ decreases monotonically tends to zero.
On the other hand, for 1.30 eV where paths from $7p$ and $8p$
interfere with each other, $\delta_{ex}$ first decreases to a
negative value before increasing again to zero. The extra phase
shift difference $\delta_{ex}$ exhibits even more peculiar behavior
at 1.22 and 1.32 eV, where the photoelectron yield strongly
oscillates with delay (see Fig.
\ref{fig:photoelectron-energy-spectra-H}); plotted within the range
$[-\pi,\pi]$ (solid lines), $\delta_{ex}$ jumps at a certain delay.
Actually, if unwrapped with a modulus of $2\pi$ (dashed lines), it
increases monotonically to $2\pi$. Small kinks around $\tau=60, 100,
140\,{\rm fs}$ for $E_{\rm kin}=1.22\, {\rm eV}$ are due to slight
numerical instability stemming from vanishing $c_S$ and/or $c_D$,
whose phases become undetermined.

\begin{figure}
\centering
\includegraphics[width=83mm]{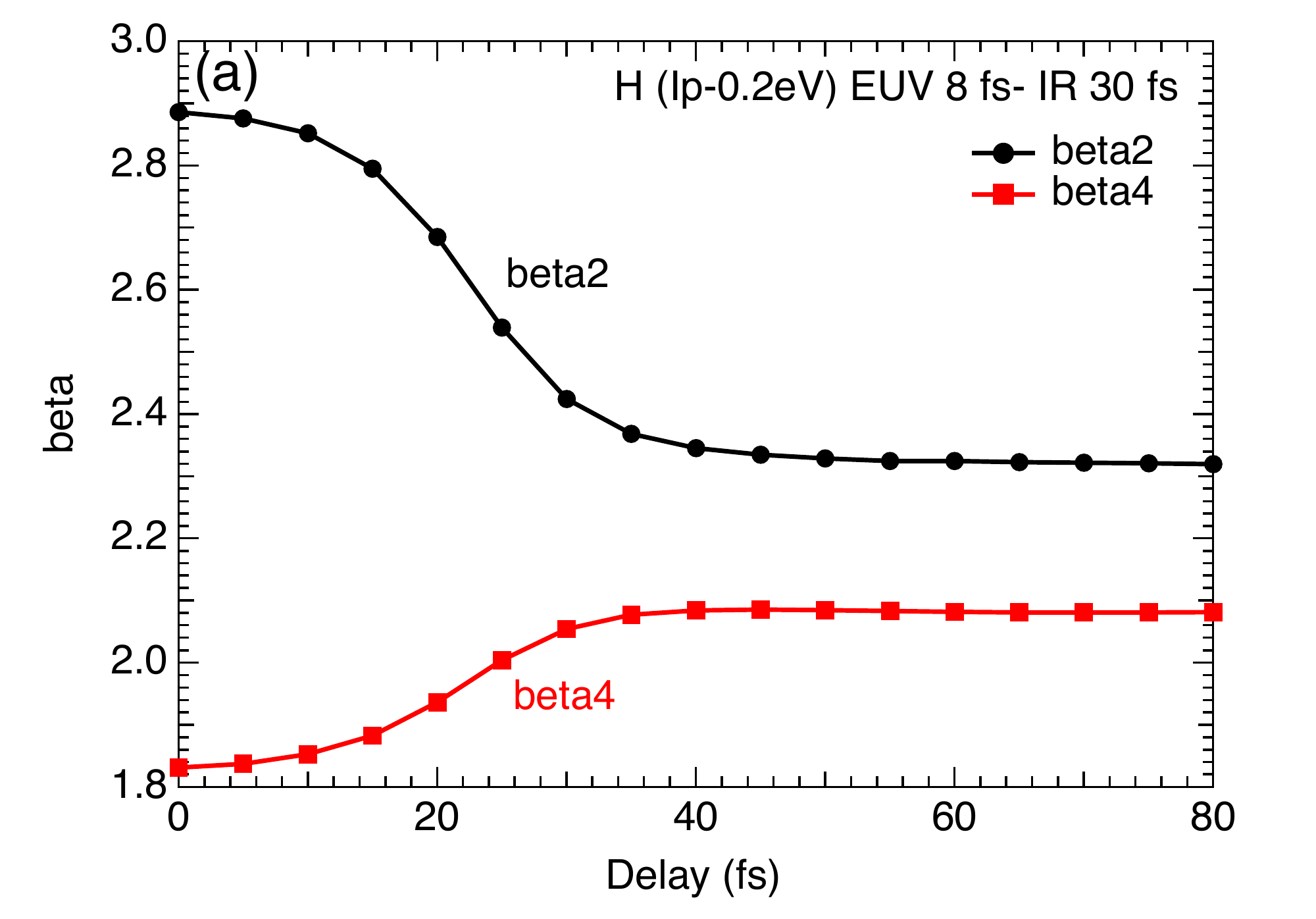}
\includegraphics[width=83mm]{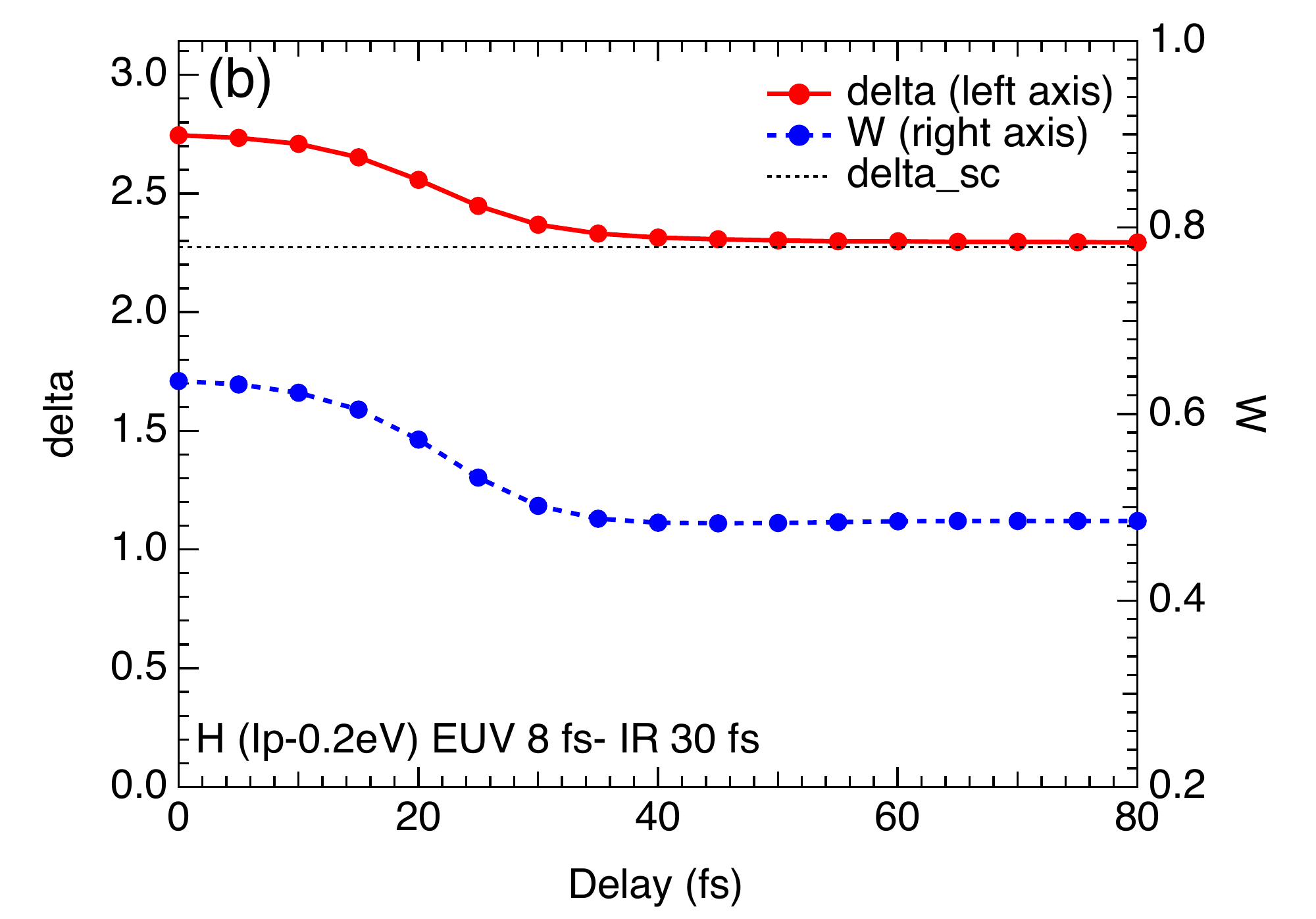}
\caption{(color online) Time-delay dependence of the energy-integrated (a)
asymmetry parameters $\beta_2$ and $\beta_4$, and (b) the relative phase $\delta$ (left axis) and amplitude
ratio $W$ (right axis) in TPSI of H atoms. Thin dashed curve: the scattering phase shift difference
$\delta_{sc} = 2.274$ (left axis).}
\label{fig:energy-integrated-H}
\end{figure}

Finally, we show the delay-dependence of photoelectron-energy
integrated asymmetry parameters $\beta_2$ and $\beta_4$ as well as
the amplitude ratio $W$ and relative phase $\delta$ in Fig.\
\ref{fig:energy-integrated-H}. As expected, all of them vary with
delay and tend to constant values. In particular, $\delta$
asymptotically tends to the scattering phase shift difference
($\delta_{sc} = 2.274$), or equivalently, $\delta_{ex}
(=\delta-\delta_{sc})$ tends to zero.

\subsection{Two-photon ionization of noble gas atoms}\label{sec:results-other}

\subsubsection{He atom}\label{sec:results-He}

\begin{figure}
\centering
\includegraphics[width=83mm]{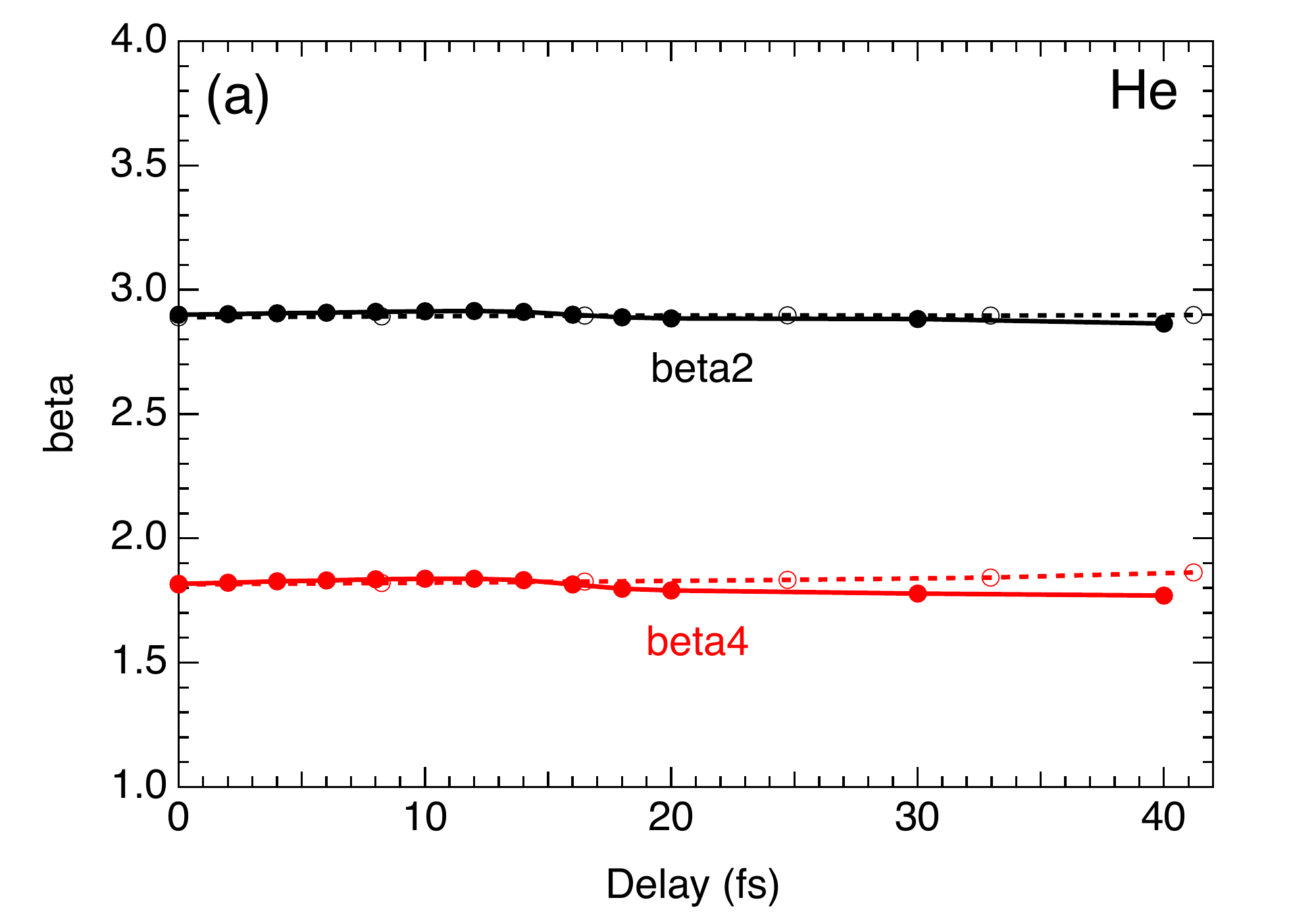}
\includegraphics[width=83mm]{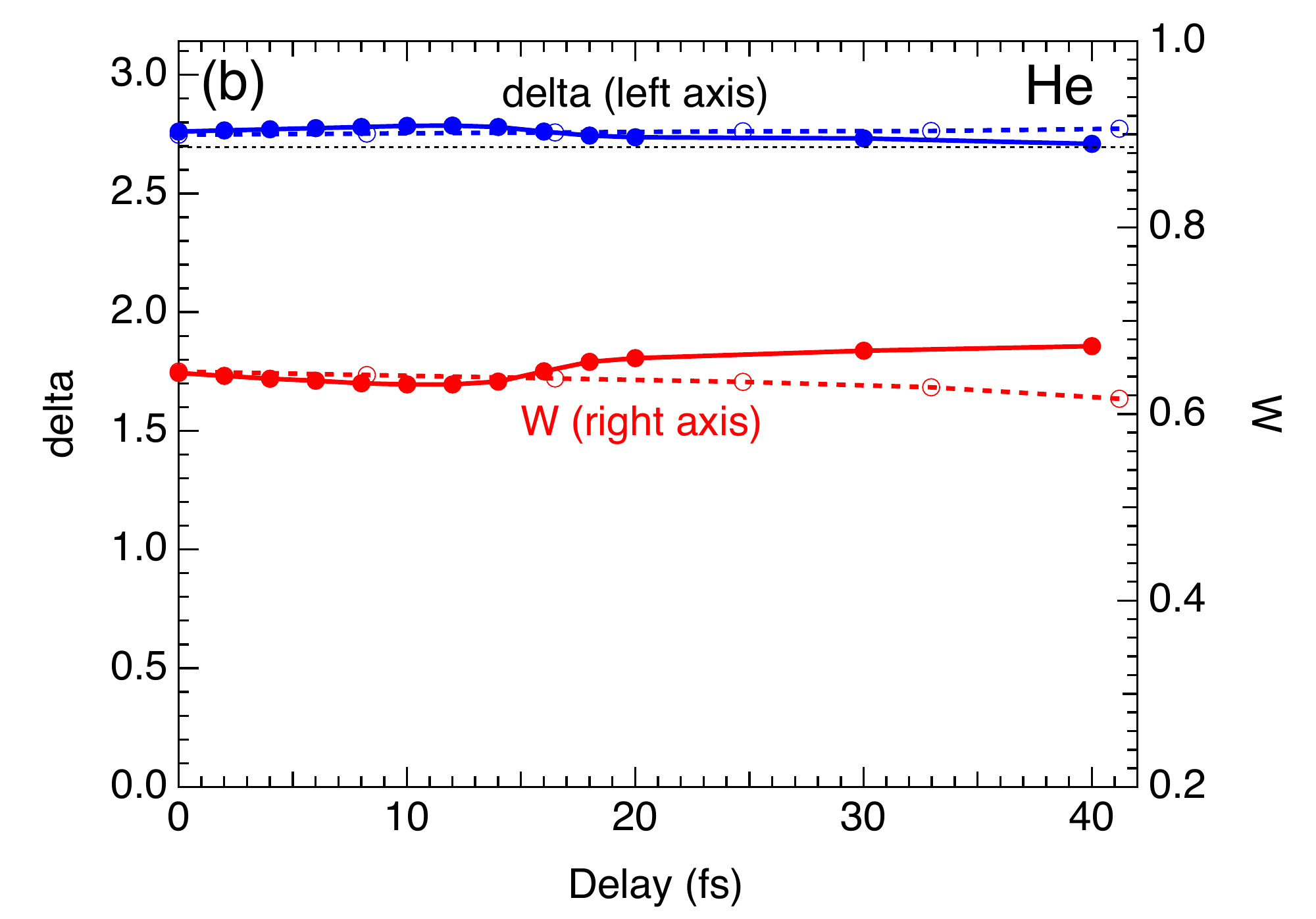}
\caption{(color online) Time-delay dependence of the energy-integrated (a)
asymmetry parameters $\beta_2$ and $\beta_4$, and (b) the relative phase $\delta$ (left axis) and amplitude
ratio $W$ (right axis) in TPSI of He atoms. Thick solid curves: two-electron TDSE simulations. Thick dashed curves: single-active-electron TDSE simulations. Thin dashed curve in (b): the scattering phase shift difference $\delta_{sc} = 2.696$ (left axis) \cite{Gien2002JPB}.}\label{fig5}
\end{figure}

The case of He is of special interest since several measurements
of the PADs from two-color TPSI of He have been reported
\cite{Guyetand08,Haber09b,Haber10,Haber11,Keeffe13,Mondal13}.
Moreover, in Refs. \cite{Guyetand08,Haber09b,Haber10} a dependence
of the PADs on the time delay between the pulses were investigated.
In Ref. \cite{Guyetand08}, however, the time-delay dependence was
studied on the attosecond scale and is connected with the relative
phase of the EUV and IR pulses which is outside the scope of our
investigation. On the other hand, in Refs. \cite{Haber09b,Haber10}
two-color TPSI of He atom in both the below- and above-threshold cases was
studied and no time-delay dependence of the PADs on the femtosecond
scale was detected within experimental errors. At first sight this
result contradicts to our main thesis that the PADs should depend on
the delay between EUV and IR pulses. To clarify this situation we
performed accurate two-electron TDSE calculations for the EUV
photon energy which is 0.2 eV below the ionization threshold. Due to limitation of
the computation time we made calculations for an IR
pulse duration of 10 fs with a peak EUV intensity of $10^{10}\,{\rm W/cm}^2$ and all the other parameters indicated in
Sec.\ \ref{sec:TDSE}. The results of calculations are shown in Fig.
\ref{fig5} as solid curves. One sees that indeed the asymmetry
parameters $\beta_2$ and $\beta_4$ as well as the amplitude ratio $W$ and the relative phase $\delta$ between the $S$ and $D$ partial waves are practically independent of
the time-delay, in agreement with the experimental reports \cite{Haber09b,Haber10}.
The value of $\delta$ is found to be close to the scattering phase shift difference $\delta_{sc}=2.696$ \cite{Gien2002JPB}.

We have also made
calculations for the same parameters within the single-active
electron approximation using the same TDSE code as for Ne and Ar. In
this case the effective single-electron potential was chosen as a
screened Coulomb potential with a polarization term:
\begin{equation}
U(r)= -\frac{2}{r}-\frac{1}{r}\left( e^{-4r} - 1\right) -2 e^{-4r} -
\frac{9}{32(r^2+1.2)^2}.
\end{equation}
The results shown in Fig. \ref{fig5} (dashed curves) are in
good agreement with a more elaborate calculations with the
two-electron TDSE. Moreover, we made single-electron calculations
for a longer IR pulse of 30 fs and have found that the beta
parameters for He are practically independent of the IR pulse
duration.
Thus we have proved that in He case, in agreement with the
experiment \cite{Haber09b,Haber10}, the PADs are practically
independent of the time delay between the pulses. The He
atom indeed represents a special case in which the PAD barely varies
with delay, accidentally, for the particular combination of photon
energies used.

In order to investigate this interesting case further we have
calculated the electron spectra for different time delays using
single active electron approximation. The spectra integrated over
emission angle are shown in Fig. \ref{fig7}. They are marked by the
numbers which indicate different relative position of the maxima of
the EUV and IR pulses as shown in Fig. \ref{fig6}. Curve 1
corresponds to a complete overlap of the pulses where their maxima
coincide. Curve 9 corresponds to another extreme case where the
pulses are separated, the EUV pulse acting first to the atom.

One can see from Fig. \ref{fig7} that the photoelectron spectrum
strongly varies with the time-delay, in striking contrast to $\beta$ parameters.
Its shape, position of the main
maximum and its width depend on the delay, reflecting the interplay
between resonant and non-resonant mechanism of ionization. The
variations of the spectra are qualitatively similar to the case of
hydrogen atom (Fig. \ref{fig:photoelectron-energy-spectra-H-2}).
When the delay is zero (line 1), both resonant and non-resonant
transitions contribute, (see Sec.\ \ref{sec:theor}), the spectrum is broad
with the maximum at $\hbar\omega_L+E_{ex}=1.35\,{\rm eV}$. Its width is mainly determined by that (0.23 eV) of the shorter EUV pulse (8 fs). In the other extreme case of non-overlapping pulses
(curve 9) particular Rydberg states (presumably mainly $1s7p$ and $1s8p$ states \cite{NIST08}) are resonantly excited by the EUV pulse, which are then ionized by the IR pulse. The main maximum is red
shifted since the lower Rydberg states are predominantly populated.
The width of the peak is smaller since it is now determined mainly
by that (0.06 eV) of the longer IR pulse (30 fs). Small maximum on the
left side of the main peak corresponds to ionization through
excitation of the $1s6p$ Rydberg state. In the intermediate cases of
partial overlap of the pulses one observes gradual transition to the
pure resonant case with interference of ionization paths via $1s6p$, $1s7p$, and $1s8p$ states.

The evergy-resolved angular distributions calculated at different parts of the
spectrum are practically the same and do not change in spite of the
variation of the spectrum, which leads to a practical independence of
the $\beta$ parameters from the pulse overlap.

\subsubsection{Ne and Ar atoms}\label{sec:results-Ne-Ar}

\begin{figure}
\centering
\includegraphics[width=83mm]{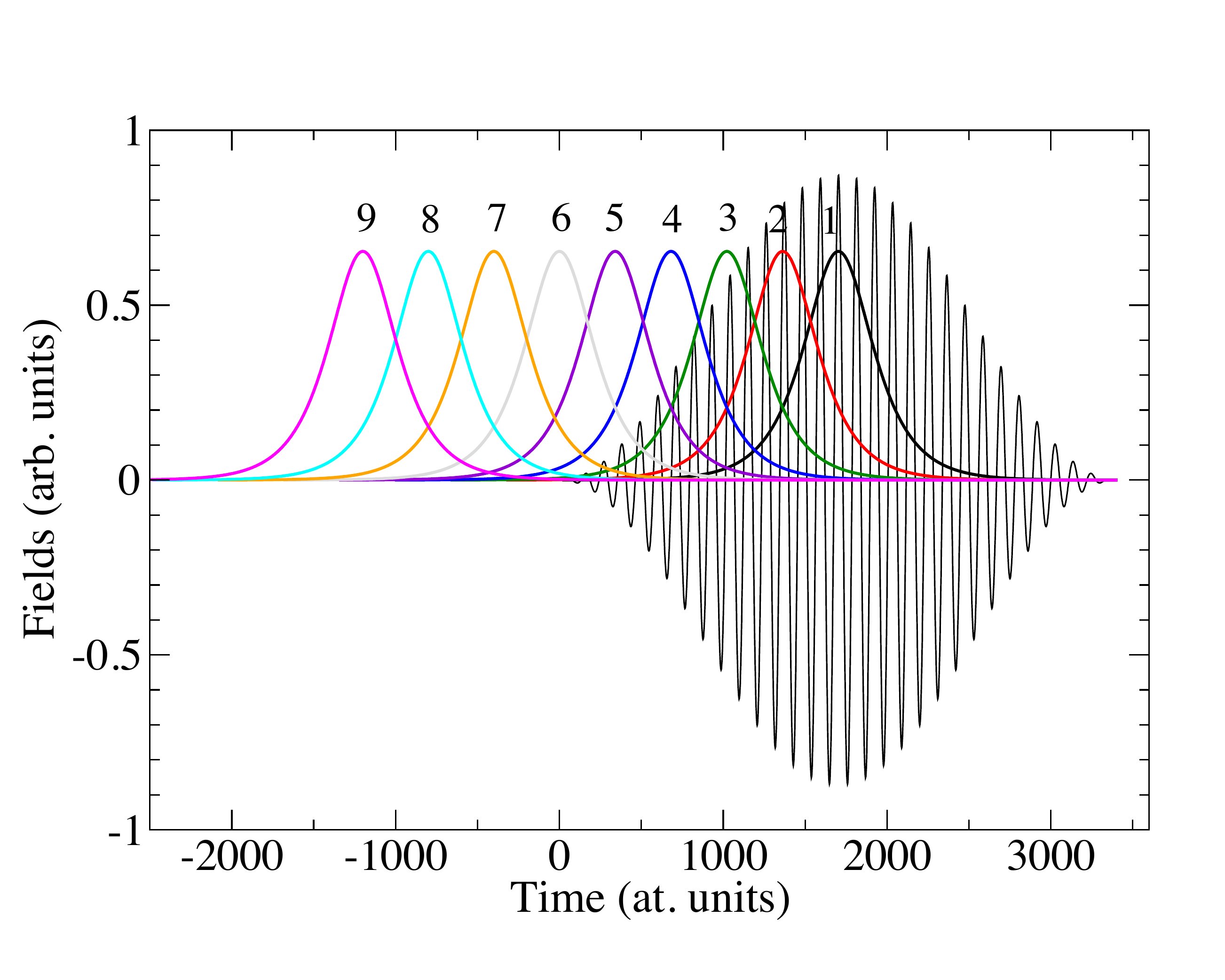}
\caption{(color online) The electric field of the 30 fs IR pulse (thin black curve)
and the envelopes of the electric field of the EUV pulses (thick
colored curves) in arbitrary units for different time delays.
Numbers indicate the following delays between EUV and IR pulses: 1 -
0 fs(complete overlap); 2 - 8.2 fs; 3 - 16.5  fs; 4 -24.7 fs; 5 - 33
fs; 6 - 41.2 fs; 7 - 50.9 fs; 8 - 60.6 fs; 9 - 70.2 fs (fully
separated pulses)}\label{fig6}
\end{figure}
\begin{figure}
\centering
\includegraphics[width=83mm]{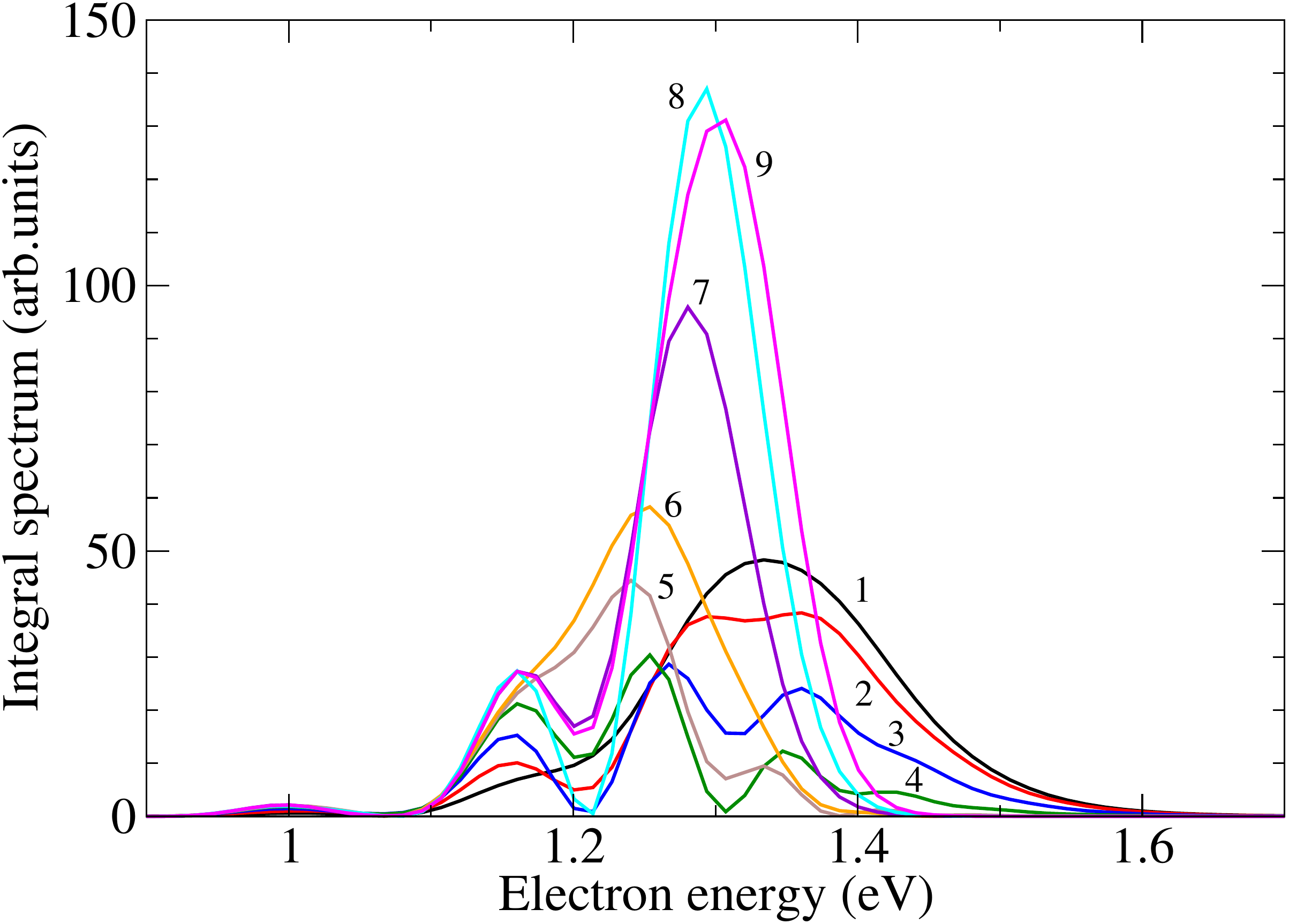}
\caption{(color online) The angle-integrated electron spectrum for TPSI of He for
varies time delays shown in Fig. \ref{fig6}. The numbers at the
curves correspond to the numbers which mark different positions of
the EUV peak in Fig. \ref{fig6}.}\label{fig7}
\end{figure}

\begin{figure}
\centering
\includegraphics[width=83mm]{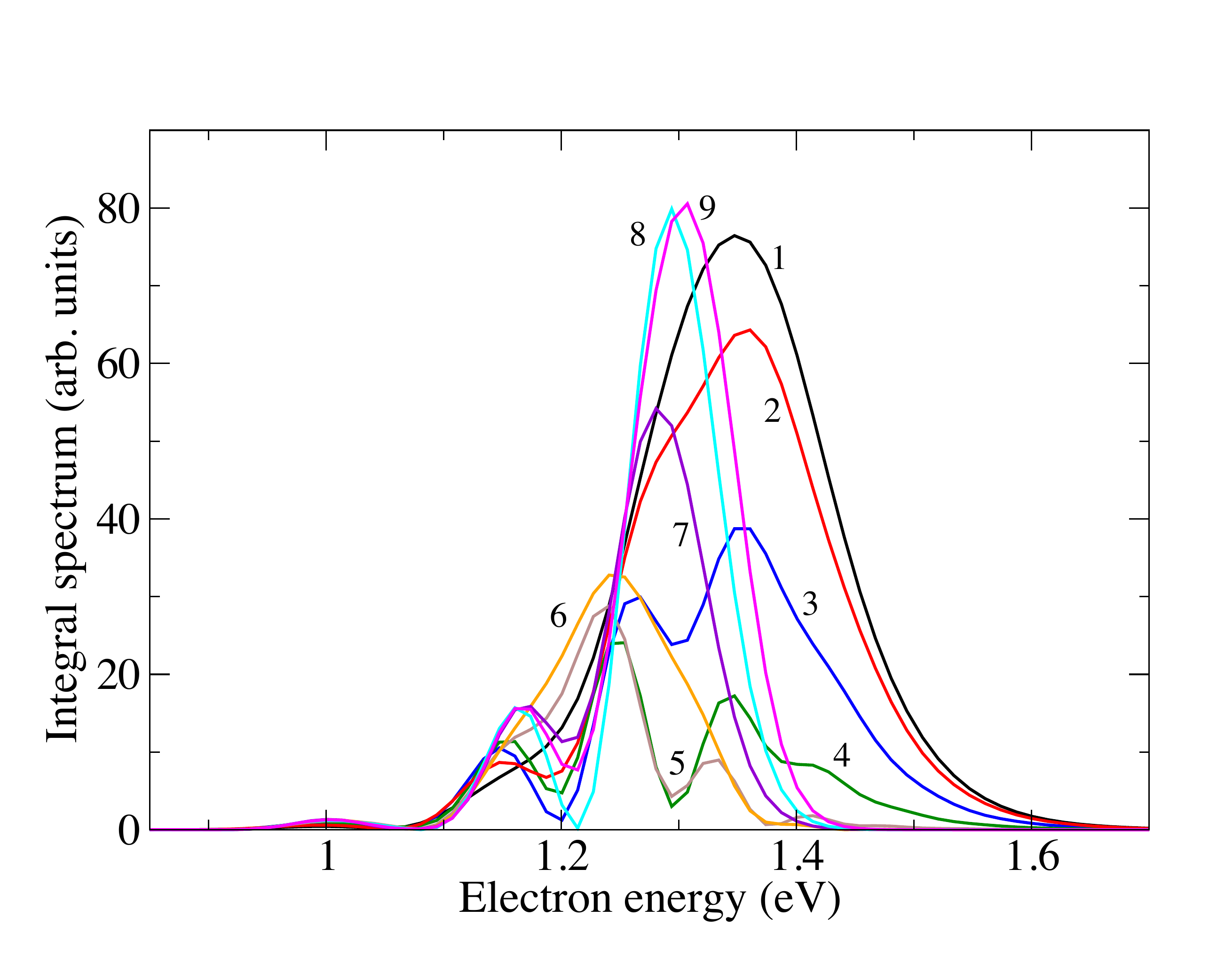}
\caption{(color online) The same as in Fig. \ref{fig7} but for $2p$ ionization of Ne
atoms.}\label{fig8}
\end{figure}

In this subsubsection we present the simulation results for Ne and Ar atoms. In both cases the EUV photon energy was
chosen to be 0.2 eV below the corresponding ionization thresholds. 
In such a case, a group
of Rydberg states is excited by the EUV pulse, which is then ionized by an IR photon. For
chosen energy of IR photon (1.55 eV) one can expect a group of
photoelectrons with the energy about 1.35 eV. In Fig. \ref{fig8} we
show the angle-integrated spectra of photoelectrons from Ne
calculated for different time delays between EUV and IR pulses from
complete overlap of the pulses (curve 1) to fully separated pulses
(curve 9). The numbers on the curves correspond to the delays
displayed in Fig. \ref{fig6}. As in the case of H and He, the shape of the
spectrum and its width strongly depends on the delay. It is mainly
determined by the interplay of the resonant and non-resonant
contributions to the ionization as discussed above.

\begin{figure}
\centering
\includegraphics[width=83mm]{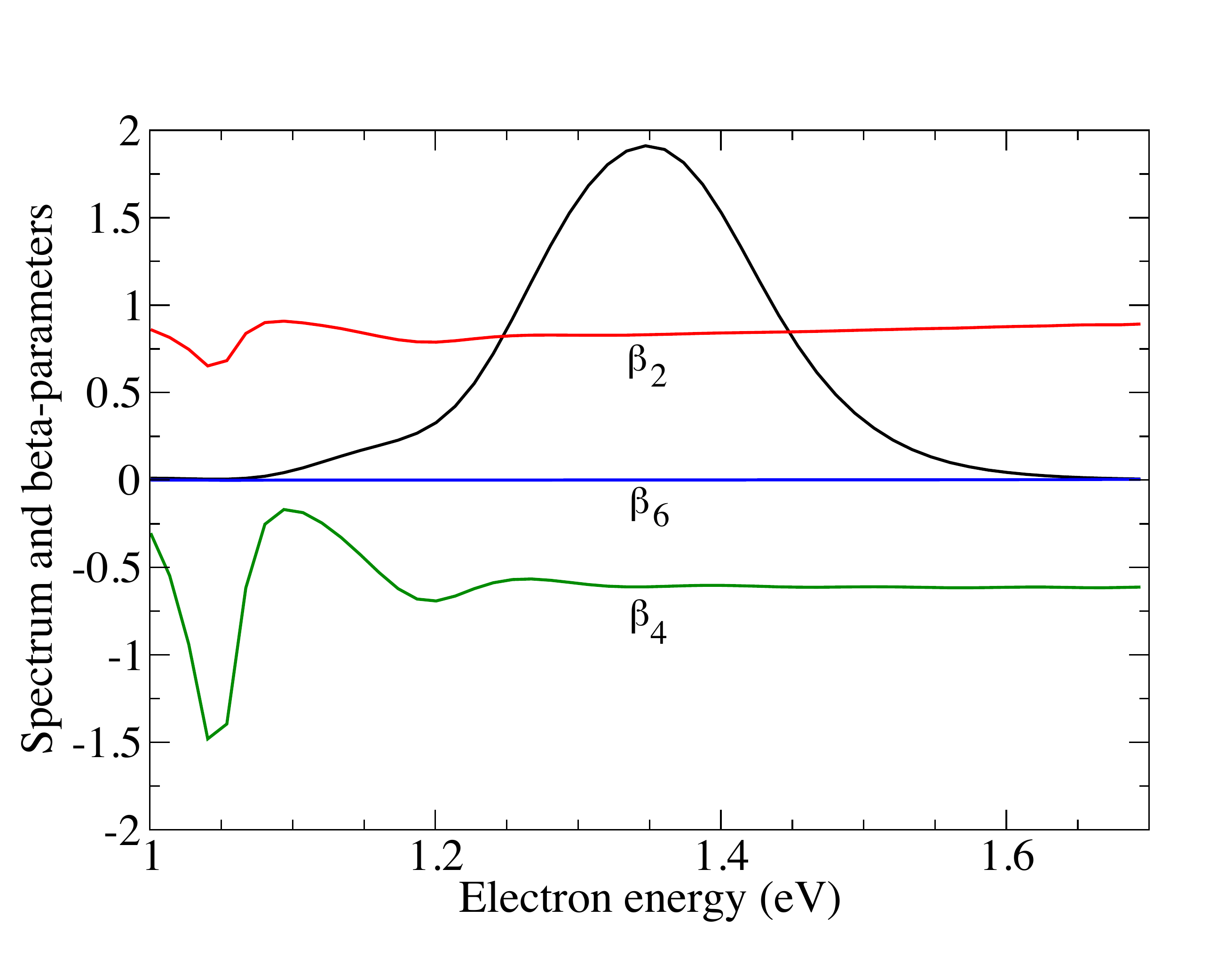}
\caption{(color online) The electron spectra (in arbitrary units) and evolution of
the asymmetry parameters across the resonance for zero time delay
between EUV and IR pulses calculated for Ne atom.}\label{fig9}
\end{figure}

For all delays we have also calculated the energy dependence of the
asymmetry parameters $\beta_2, \beta_4$ and $\beta_6$, where $\beta_6$ is the next coefficient in the expansion of the PAD in terms of Legendre polynomials. In all cases
the latter parameter is at least two orders of magnitude smaller
than the first two. This confirms that at the chosen IR intensity of
$10^{10}$ W/cm$^2$ only one IR photon is absorbed. Together with the
EUV excitation it gives two-photon ionization with angular
distribution of photoelectrons described by Eq.
(\ref{eq:pad}). As an example in Fig. \ref{fig9} we show the
photoelectron spectrum from Ne and $\beta$ parameters as functions of
photoelectron energy for the case of complete overlap of the EUV and
IR pulses (case 1 in Fig. \ref{fig6}). Interestingly, the parameters
$\beta_2$ and $\beta_4$ are practically constant in the region of
maximum, changing their value only when the cross section is small, which is consistent with the hydrogen case (see Fig.\ \ref{fig:delta-tau-0}).
Similar behavior is observed for all other delays.

\begin{figure}
\centering
\includegraphics[width=83mm]{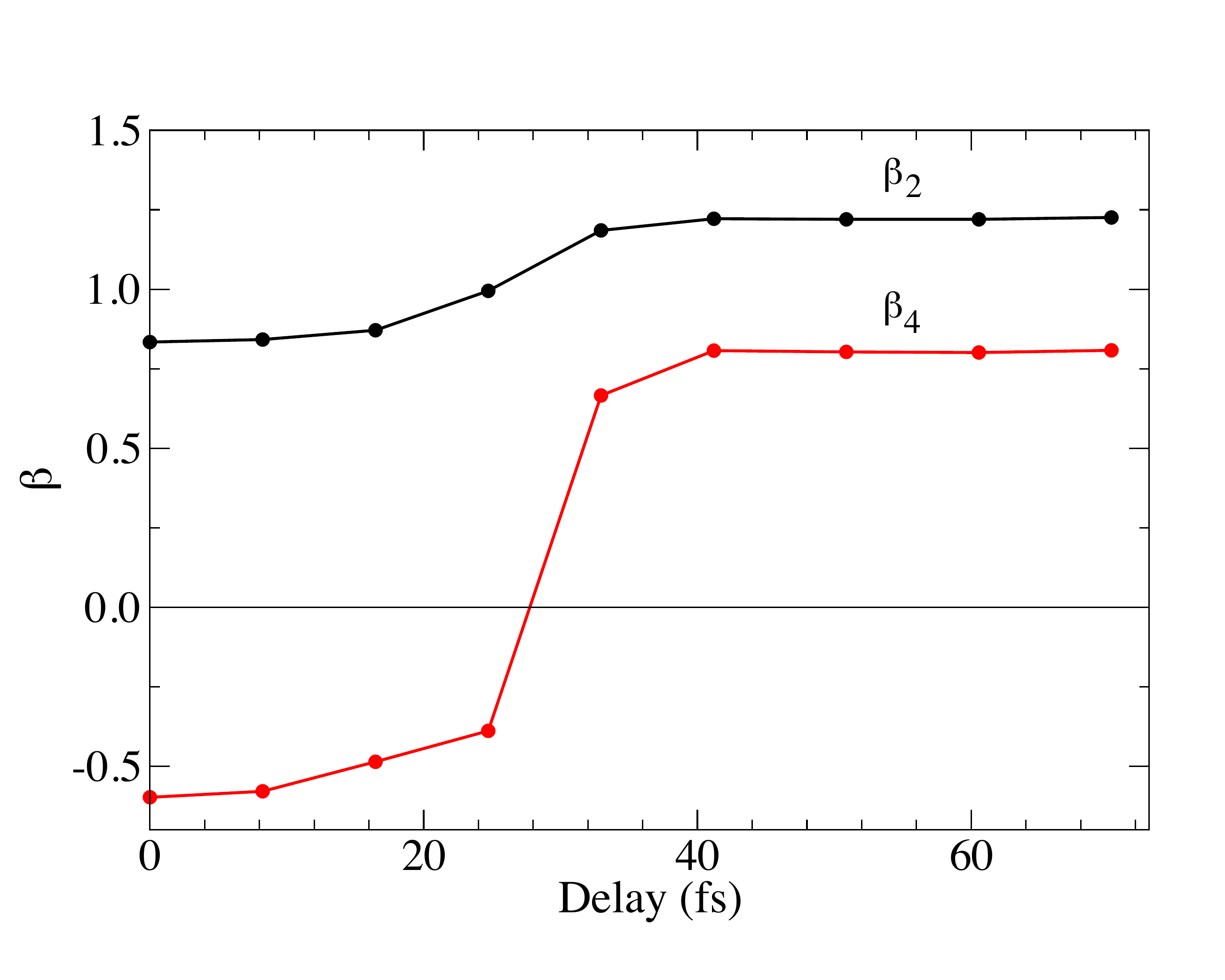}
\caption{(color online) The calculated dependence of asymmetry parameters $\beta_2$
and $\beta_4$ on delay between EUV and IR pulses for the case of Ne
ionization at EUV photon energy $-0.2$ eV below threshold. The
parameters are shown for the angular distribution integrated over
the peak. The points are connected by straight lines to guide the eye.}\label{fig10}
\end{figure}

In Fig. \ref{fig10} we show the calculated $\beta_2$ and $\beta_4$
parameters for the angular distribution integrated over the peak, as
it is usually measured in real experiments. The parameters are shown
as functions of time delay. One sees that both parameters are
changing considerably with the delay. The $\beta_4$ even changes its
sign. This behavior was predicted theoretically and confirmed by
experiment in our recent publication \cite{Mondal14}.

\begin{figure}
\centering
\includegraphics[width=83mm]{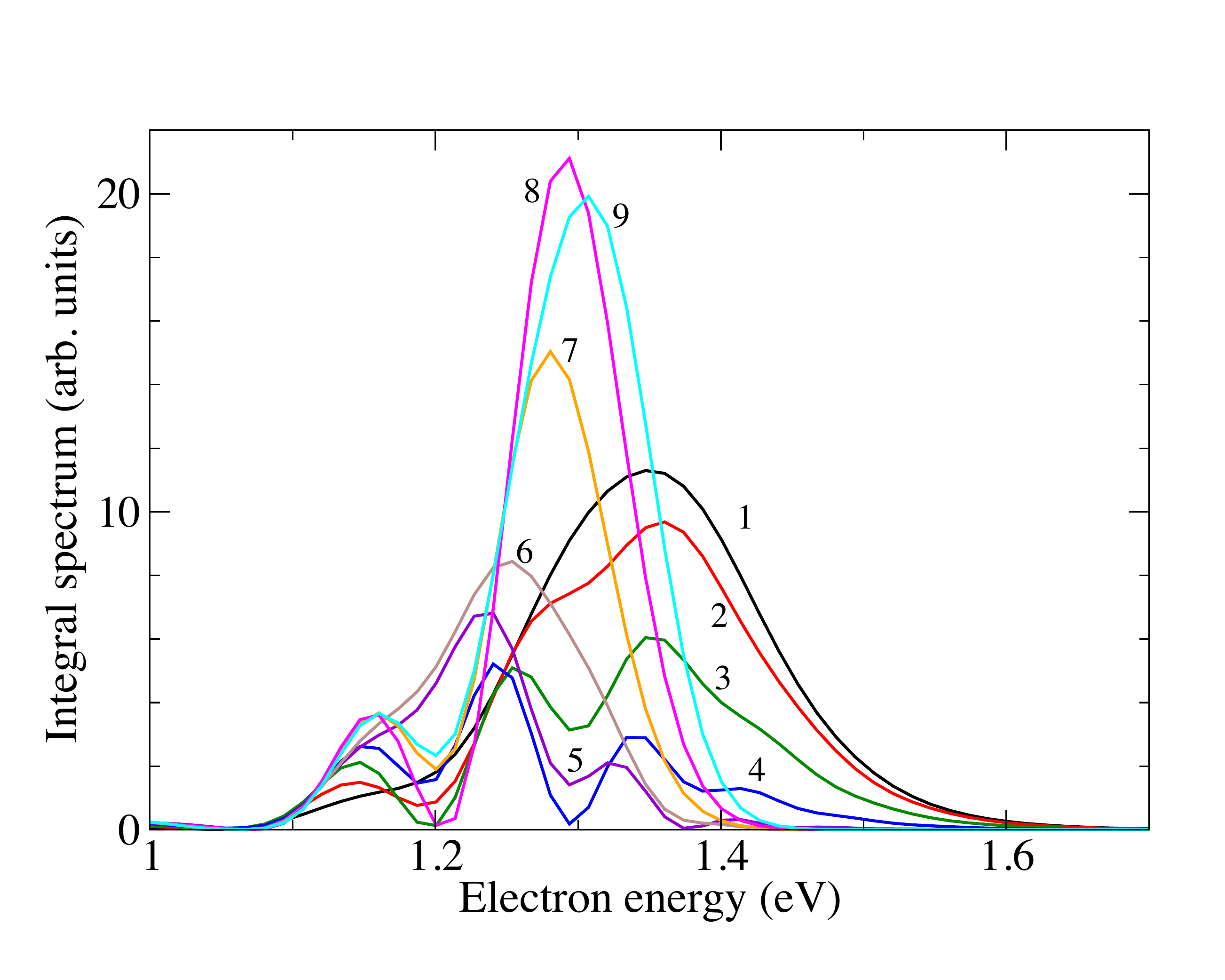}
\caption{(color online) Photoelectron spectra integrated over emission angle for
different time-delays indicated in Fig. \ref{fig6}, calculated for
3p ionization of Ar.}\label{fig11}
\end{figure}

Similar calculations have been done for Ar. The calculated
photoelectron spectra integrated over the emission angle are
presented in Fig. \ref{fig11} for several delays between pulses.
Qualitatively the spectra and their variation with the time-delay
are similar to the cases of H (Fig. \ref{fig:photoelectron-energy-spectra-H-2}), He (Fig. \ref{fig7}) and
Ne (Fig. \ref{fig8}). This is natural since the properties of the
Rydberg states close to the threshold depend  only weakly on the
properties of the core.

\begin{figure}
\centering
\includegraphics[width=83mm]{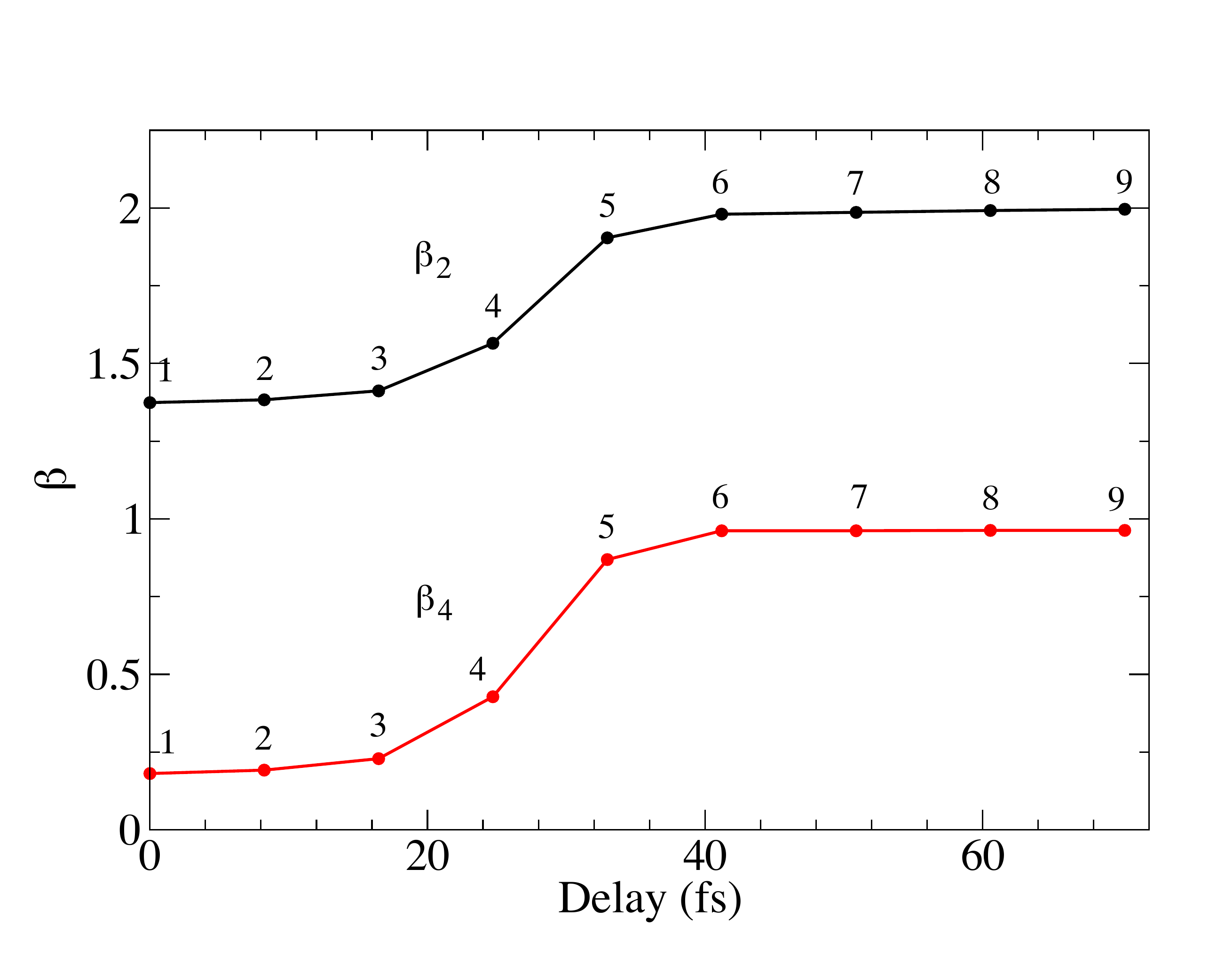}
\caption{(color online) Photoelectron angular distribution parameters $\beta_2$ and
$\beta_4$ for photoionization of Ar atom as functions of time-delay.
Numbers indicate the particular delays shown in Fig. \ref{fig6}. The
points are connected by straight lines to guide the eye.
}\label{fig12}
\end{figure}

Figure \ref{fig12} shows the values of $\beta_2$ and $\beta_4$
for the case of Ar, calculated for various time-delays. Similar to the Ne case the
asymmetry parameters notably depend on the delay. Interestingly, in
the Ar case the $\beta_4$ parameter does not change its sign unlike in
the case of Ne. This difference is possibly explained by different $s$ and $d$ excitation by the EUV pulse in Ne and Ar \cite{KM72}.


According to our calculations the variation of the $\beta_2$ and
$\beta_4$ parameters with the time delay is much more pronounced for
Ne and Ar than for H and He atoms. Presumably, this is related to
the fact that in Ne and Ar $p$-electron is ionized. In this case the
PAD in two-photon ionization is defined mainly by the contribution
of $P$ and $F$ partial wave packets which can give more space for
beta variations.  In particular, the $\beta_4$ parameter in
$s$-ionization depends only on the ratio W of $S$ and $D$ amplitudes
(see Eqs. (\ref{eq:beta2and4})), while in $p$-ionization it depends
on both the amplitude ratio and relative phase of $P$ and $F$ partial waves, which may be more sensitive to the contribution of resonant
and non-resonant pathways.

\section{Conclusions}\label{sec:conclusions}

We have theoretically investigated the PADs for
two-color (EUV+IR) TPSI of H, He, Ne, and Ar atoms with EUV
excitation slightly below the ionization threshold. The PADs for EUV+IR TPSI have recently been experimentally measured with modern EUV FEL and high-harmonic sources. We have shown
that the photoelectron energy spectra as well as anisotropy parameters $\beta_2$ and $\beta_4$ strongly depend on the time delay between the EUV and
IR pulses, except for $\beta$ values for the case of He. This dependence is associated with the contributions of the resonant and nonresonant pathways of
ionization, changing with the pulse delay, which implies that investigations of the time-delay dependence
of the PADs in TPSI make it possible to study the fundamental
problem of the interplay of resonant and nonresonant
processes in photoionization. Our results indicate that the
variation of PADs with the time delay is more pronounced for
ionization of $p$-shell electrons (Ne and Ar) than for $s$-shell
electrons (H and He). Surprisingly, the anisotropy parameters barely changes with delay for the case of He for the present combination of photon energies. This explains why the delay dependence was not detected in \cite{Haber09b}.

\begin{acknowledgments}

KLI gratefully acknowledges support by KAKENHI (Grants No. 23656043,
No. 25286064, and No. 26600111), the Photon Frontier Network program
of MEXT (Japan), the Center of Innovation Program from JST (Japan),
and the Cooperative Research Program of ``Network Joint Research
Center for Materials and Devices'' (Japan). KLI also thanks Mr. Y. Futakuchi and Ms. S. Watanabe 
for their help in Fig.\ \ref{fig:delta-delay} and reference preparation, respectively. NMK acknowledges
financial support from the programme 'Physics with Accelerators and
Reactors in West Europe' of the Russian Ministry of Education and
Science. He also acknowledges hospitality and financial support from
XFEL (Hamburg). KU is grateful to the Ministry of Education,
Culture, Sports, Science, and Technology of Japan for supports via
X-ray Free Electron Laser Utilization Research Project, the X-ray
Free Electron Laser Priority Strategy Program, and Management
Expenses Grants for National Universities Corporations.

\end{acknowledgments}

\end{document}